\documentclass{article}
\usepackage[margin=2cm]{geometry}
\usepackage[T1]{fontenc}
\usepackage[utf8]{inputenc}
\usepackage{amsmath,amssymb,amsthm}
\usepackage[shortlabels]{enumitem}
\usepackage{color,url}
\usepackage{hyperref}
\usepackage[capitalise]{cleveref}
\usepackage{xspace}
\usepackage{todonotes}
\usepackage{thmtools}
\usepackage{thmtools}
\makeatletter
\@ifundefined{newcounteralias}{}{%
  \renewcommand\thmt@autorefsetup{%
    \@xa\def\csname\thmt@envname autorefname\@xa\endcsname
      \@xa{\thmt@thmname}%
  }%
}
\makeatother

\usepackage{comment}
\usepackage{xcolor}
\usepackage{colortbl}
\usepackage{authblk}
\usepackage{libertine}
\usepackage{cite}
\usetikzlibrary{decorations.pathreplacing,calligraphy}

\theoremstyle{plain}
\newtheorem{theorem}{Theorem}[section]
\newtheorem{claim}{Claim}[theorem]
\newtheorem{lemma}[theorem]{Lemma}
\newtheorem{proposition}[theorem]{Proposition}

\newenvironment{claimproof}{\noindent\textit{Proof of Claim \theclaim:}}{\hfill$\lrcorner$\medskip}

\crefname{claim}{Claim}{Claims}

\newcommand\blfootnote[1]{%
  \begingroup
  \renewcommand\thefootnote{}\footnote{#1}%
  \addtocounter{footnote}{-1}%
  \endgroup
}

\renewcommand{\phi}{\varphi}

\renewcommand{\epsilon}{\varepsilon}

\newcommand{\eps}{\epsilon}
\newcommand{\Oh}{\mathcal{O}}

\newcommand{\cF}{\mathcal{F}}
\newcommand{\cG}{\mathcal{G}}

\newcommand{\cQ}{\mathcal{Q}}
\newcommand{\cR}{\mathcal{R}}

\newcommand{\cY}{\mathcal{Y}}

\newcommand{\tw}{\mathsf{tw}}

\newcommand{\col}[1]{{#1}-\textsc{Coloring}\xspace}
\newcommand{\lcol}[1]{\textsc{List} {#1}-\textsc{Coloring}\xspace}

\newcommand{\NP}{\textsf{NP}\xspace}

\newcommand{\up}{\mathsf{up}}
\newcommand{\down}{\mathsf{down}}
\newcommand{\Up}{\mathsf{Up}}
\newcommand{\Down}{\mathsf{Down}}
\newcommand{\Top}{\mathsf{Top}}
\newcommand{\Bottom}{\mathsf{Bottom}}

\newcommand{\abab}{
\begin{tikzpicture}[every node/.style={fill=black, circle, inner sep=0.5pt, outer sep = 0}, scale=0.2,baseline= -0.5mm]
    \node (a1) at (0,0) {};
    \node (b1) at (1,0) {};
    \node (a2) at (2,0) {};    
    \node (b2) at (3,0) {};    
    \path[draw] (a1) to[bend left = 60] (a2);
    \path[draw] (b1) to[bend left = 60] (b2);
\end{tikzpicture}
}

\newcommand{\ababk}{
\begin{tikzpicture}[every node/.style={fill=black, circle, inner sep=0.5pt, outer sep = 0}, scale=0.2,baseline= -0.5mm]
    \node[fill=none] (k0) at (-1,0) {\footnotesize{$_{(k)}$}};   
    \node (a1) at (0,0) {};
    \node[fill=none] (k1) at (1,0) {\footnotesize{$_{(k)}$}};   
    \node (b1) at (2,0) {};
    \node[fill=none] (k2) at (3,0) {\footnotesize{$_{(k)}$}};   
    \node (a2) at (4,0) {};    
    \node[fill=none] (k3) at (5,0) {\footnotesize{$_{(k)}$}};   
    \node (b2) at (6,0) {}; 
    \node[fill=none] (k4) at (7,0) {\footnotesize{$_{(k)}$}};      
    \path[draw] (a1) to[bend left = 70] (a2);
    \path[draw] (b1) to[bend left = 70] (b2);
\end{tikzpicture}
}

\newcommand{\aabb}{
\begin{tikzpicture}[every node/.style={fill=black, circle, inner sep=0.5pt, outer sep = 0}, scale=0.2,baseline= -0.5mm]
    \node (a1) at (0,0) {};
    \node (a2) at (1,0) {};
    \node (b1) at (2,0) {};    
    \node (b2) at (3,0) {};    
    \path[draw] (a1) to[bend left = 60] (a2);
    \path[draw] (b1) to[bend left = 60] (b2);
\end{tikzpicture}
}

\newcommand{\abba}{
\begin{tikzpicture}[every node/.style={fill=black, circle, inner sep=0.5pt, outer sep = 0}, scale=0.2,baseline= -0.5mm]
    \node (a1) at (0,0) {};
    \node (b1) at (1,0) {};
    \node (b2) at (2,0) {};    
    \node (a2) at (3,0) {};    
    \path[draw] (a1) to[bend left = 70] (a2);
    \path[draw] (b1) to[bend left = 45] (b2);
\end{tikzpicture}
}

\newcommand{\abbak}{
\begin{tikzpicture}[every node/.style={fill=black, circle, inner sep=0.5pt, outer sep = 0}, scale=0.2,baseline= -0.5mm]
    \node[fill=none] (k0) at (-1,0) {\footnotesize{$_{(k)}$}};  
    \node (a1) at (0,0) {};
    \node[fill=none] (k1) at (1,0) {\footnotesize{$_{(k)}$}};  
    \node (b1) at (2,0) {};
    \node (b2) at (3,0) {};    
    \node[fill=none] (k2) at (4,0) {\footnotesize{$_{(k)}$}};  
    \node (a2) at (5,0) {};    
    \node[fill=none] (k3) at (6,0) {\footnotesize{$_{(k)}$}};  
    \path[draw] (a1) to[bend left = 70] (a2);
    \path[draw] (b1) to[bend left = 45] (b2);
\end{tikzpicture}
}

\newcommand{\akbbka}{
\begin{tikzpicture}[every node/.style={fill=black, circle, inner sep=0.5pt, outer sep = 0}, scale=0.2,baseline= -0.5mm] 
    \node (a1) at (0,0) {};
    \node[fill=none] (k1) at (1,0) {\footnotesize{$_{(k)}$}};  
    \node (b1) at (2,0) {};
    \node (b2) at (3,0) {};    
    \node[fill=none] (k2) at (4,0) {\footnotesize{$_{(k)}$}};  
    \node (a2) at (5,0) {};
    \path[draw] (a1) to[bend left = 70] (a2);
    \path[draw] (b1) to[bend left = 45] (b2);
\end{tikzpicture}
}

\newcommand{\jj}{
\begin{tikzpicture}[every node/.style={fill=black, circle, inner sep=0.5pt, outer sep = 0}, scale=0.2]
    \node (x) at (0,0) {};
    \node (y) at (1,0) {};
    \node (z) at (2,0) {};    
    \path[draw] (x) to[bend left = 45] (y);
    \path[draw] (x) to[bend left = 60] (z);
\end{tikzpicture}
}

\newcommand{\jjp}{
\begin{tikzpicture}[every node/.style={fill=black, circle, inner sep=0.5pt, outer sep = 0}, scale=0.2]
    \node (x) at (0,0) {};
    \node (y) at (1,0) {};
    \node (z) at (2,0) {};    
    \node (q) at (3,0) {};    
    \path[draw] (x) to[bend left = 45] (y);
    \path[draw] (x) to[bend left = 60] (z);
\end{tikzpicture}
}

\newcommand{\kjj}{
\begin{tikzpicture}[every node/.style={fill=black, circle, inner sep=0.5pt, outer sep = 0}, scale=0.2,baseline= -0.5mm]
    \node[fill=none] (pre) at (-1,0) {\footnotesize{$_{(k)}$}};   
    \node (x) at (0,0) {};
    \node (y) at (1,0) {};
    \node (z) at (2,0) {};        
    \path[draw] (x) to[bend left = 45] (y);
    \path[draw] (x) to[bend left = 60] (z);
\end{tikzpicture}
}

\newcommand{\lkj}{_{(k)}\jj_{(\ell)}}

\renewcommand{\phi}{\varphi}

\title{List coloring in hereditary classes of ordered graphs, revisited}
\author[1]{Paweł Rafał Bieliński}
\author[1]{Michał Dębski}
\author[2]{Marta Piecyk}
\author[1]{Pawe{\l} Rz\k{a}\.zewski}
\author[1]{Jakub Żuchowski}

\affil[1]{Warsaw University of Technology}
\affil[2]{CISPA Helmholtz Center for Information Security}
\date{}

\begin{document}
	\begin{titlepage}
	\date{}
	\maketitle

\begin{abstract}
For a fixed positive integer $k$, in the \textsc{List} $q$-\textsc{Coloring} problem, we are given a graph $G$ whose every vertex $v$ is equipped with a list $L(v)\subseteq \{1,\ldots,q\}$, and the task is to determine whether there exists a proper coloring that respects the lists $L$.

Hajebi, Li, and Spirkl [SIAM J. Discr. Math. 38 (2024)] initiated the study of the complexity of the problem for $k=3$ on \emph{$H$-free ordered graphs}, i.e., graphs given with a linear ordering of vertices that do not contain some fixed ordered graph $H$ as an induced subgraph.
This line of research was continued by Piecyk and Rzążewski [STACS 2026], who studied \textsc{List} $q$-\textsc{Coloring} for larger values of $k$, focusing mostly on the case $q=4$.

In this paper, we investigate the remaining gaps in the complexity of \textsc{List} $q$-\textsc{Coloring} on $H$-free ordered graphs.
We provide a full dichotomy, for each $q$ and $H$ classifying the problem as either quasipolynomial-time solvable, or \NP-hard.
For $q \geq 4$, we actually show a stronger dichotomy into polynomial-time solvable and \NP-hard cases.

Finally, we revisit some of the \NP-hard cases showing a subexponential-time algorithm and some (quasi)polynomial-time algorithms for the promise variant of the problem.
\end{abstract}
  \blfootnote{PRB and PRz were supported by the National Science Centre grant number 2024/54/E/ST6/00094.}

	\thispagestyle{empty}    
    \tableofcontents
\end{titlepage}
\newpage\setcounter{page}{1}

\maketitle

\newpage

\section{Introduction}

In the \textsc{List} $q$-\textsc{Coloring} problem we are given a graph $G$ and a function $L$ that assigns a subset of the set $\{1,\ldots,q\}$ to each vertex of $G$; we ask whether there exists a proper coloring of $G$ such that for each vertex $v$, the color of $v$ is in $L(v)$.
This problem generalizes standard (non-list) $q$-\textsc{Coloring}. 

A long line of research is dedicated to determine the complexity of the problem when the input graph is restricted to be $H$-free, i.e., does not contain an induced subgraph isomorphic to a fixed graph $H$. We know many values of $q$ and $H$ for which the \textsc{List-}$q$-\textsc{Coloring} problem on $H$-free graphs is \NP-hard or polynomial-time solvable~\cite{DBLP:journals/cpc/Emden-WeinertHK98,DBLP:journals/siamcomp/Holyer81a,DBLP:journals/jal/LevenG83,DBLP:journals/corr/abs-2505-00412,DBLP:journals/combinatorica/ChudnovskyHS24,DBLP:journals/siamdm/HajebiLS22,DBLP:journals/algorithmica/HoangKLSS10,DBLP:journals/iandc/GolovachPS14,DBLP:journals/algorithmica/0001GKP15,DBLP:journals/ejc/Huang16,DBLP:journals/combinatorica/BonomoCMSSZ18,DBLP:journals/algorithmica/ChudnovskyHSZ21}. The classification is particularly interesting --- and elusive --- for $q=3$. It seems that in this case the problem is polynomial if and only if $H$ is a linear forest (i.e. a forest of maximum degree at most $2$).
This belief is supported by known hardness results~\cite{DBLP:journals/cpc/Emden-WeinertHK98,DBLP:journals/siamcomp/Holyer81a,DBLP:journals/jal/LevenG83} and a quasipolynomial-time algorithm~\cite{DBLP:conf/sosa/PilipczukPR21}, and confirmed for a number of small cases~\cite{DBLP:journals/combinatorica/BonomoCMSSZ18,DBLP:journals/algorithmica/ChudnovskyHSZ21}.

Our goal is to understand the complexity of the problem on ordered graphs (where a graph is presented together with a linear ordering of its vertices). The ordering of vertices is immaterial, in a sense that it does not interact with the notion of proper vertex coloring, but it allows a finer distinction between classes of graphs that are given as inputs, i.e., a class of ordered graphs without a monotonically ordered path on $t$ vertices is much richer than a class of $P_t$-free graphs; the latter can be characterized by excluding all possible orderings of a $t$-vertex path.
We aim to obtain a complete classification of all pairs of a natural number $q$ and an ordered graph $H$ by the complexity of the \textsc{List} $q$-\textsc{Coloring} problem on $H$-free ordered graphs. This line of research was initiated by Hajebi, Li and Spirkl for $q=3$~\cite{DBLP:journals/siamdm/HajebiLS24} and continued by Piecyk and Rzążewski for $q \geq 4$~\cite{DBLP:conf/stacs/PiecykR26}.

Before we proceed to the discussion of known and new results, let us say a few words about notation. Following~\cite{DBLP:journals/corr/abs-2604-24343,DBLP:conf/stacs/PiecykR26}, we use icons to denote ordered graphs. For example, \jj denotes an ordered graph on three vertices $a,b,c$, in this order, and edges $ab,ac$.
By $(k)$ we denote a sequence of $k$ isolated vertices, so $\kjj$ is the graph $\jj$ preceded by $k$ isolated vertices.

\begin{figure}[t]
    \centering
    \includegraphics[scale=0.85]{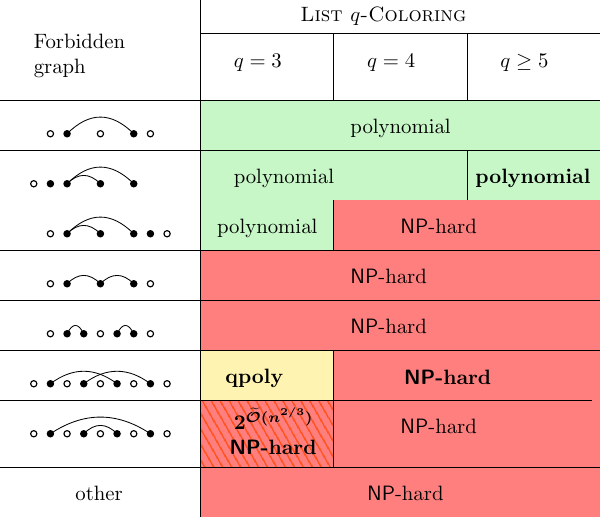}
    \caption{Known complexity results concerning \lcol{$q$} in classes of ordered graphs defined by one forbidden induced subgraph.
    Empty dots indicate any number of vertices (possibly 0).
    Results listed with bold font are shown in this paper.}
    \label{fig:summary}
\end{figure}

If $H$ is an ordered graph with at least three edges, then \textsc{List} $q$-\textsc{Coloring} on $H$-free ordered graphs is \NP-complete for all $q\geq 3$~\cite{DBLP:journals/siamdm/HajebiLS24}. It leaves seven fundamentally distinct cases for $H$ that are shown in Figure \ref{fig:summary}; each consists of one or two edges and possibly some isolated vertices (and, as it turns out, the exact number of those vertices does not make a substantial difference, as long as it is constant). 

Let us highlight the case of $\kjj$-free graphs.
As observed in~\cite{DBLP:journals/siamdm/HajebiLS24}, \jj-free ordered graphs are chordal.
Since in \textsc{List} $q$-\textsc{Coloring} we can always assume that the input graph does not contain a complete graph on $q+1$ vertices, otherwise it is clearly a no-instance, we are actually dealing with graphs of bounded treewidth, which admit a polynomial algorithm for \textsc{List} $q$-\textsc{Coloring} for every $q$.
In case of $q=3$, this result can be extended for arbitrary number of isolated vertices preceding or succeding \jj, i.e., \textsc{List} $3$-\textsc{Coloring} is polynomial-time solvable on $\lkj$-free graphs~\cite{DBLP:journals/siamdm/HajebiLS24}.
Furthermore, a polynomial-time algorithm was also provided for $q=4$ and \kjj-free ordered graphs~\cite{DBLP:conf/stacs/PiecykR26}.
Let us point out that the latter algorithm is very specific for the case of 4 colors, and cannot be (easily) extended to $q \geq 5$.

\subparagraph{Completing the dichotomy.}
As our first result, we give a polynomial-time algorithm for \lcol{$q$} in $\kjj$-free graphs, for every fixed $k$ and $q$. As the algorithm is not specific to any fixed number of color, our approach allows us to understand the structure of \kjj-free ordered graphs better.

\begin{restatable}{theorem}{thmfork}\label{thm:fork}
    For every fixed $k$ and $q$,
    \lcol{$q$} in \kjj-free ordered graphs can be solved in polynomial time.
\end{restatable}

It is in a stark contrast with the fact that for \jjp-free ordered graphs, the problem becomes \NP-complete already for $q= 4$~\cite{DBLP:conf/stacs/PiecykR26}. This shows how important the placement of isolated vertices in the forbidden subgraph can be.

Now we turn our attention to \abab.
We show that \abab-free ordered graphs are at least as rich as outerstring graphs.

\begin{restatable}{theorem}{thmababouterstring}\label{thm:abab-outerstring}
    Every outerstring graph admits an ordering that is \abab-free.
    Furthermore, given an outerstring representation of an $n$-vertex graph,
    we can find such an ordering in time polynomial in the number of vertices.
\end{restatable}

\cref{thm:abab-outerstring} combined with result of Unger~\cite{DBLP:conf/stacs/Unger88} for outerstring graphs implies the following hardness (even in the non-list setting).

\begin{restatable}{theorem}{thmababhard}\label{thm:abab-hardness}
    For every fixed $q \geq 4$, \col{$q$} is \NP-hard on \abab-free ordered graphs.
\end{restatable}

\cref{thm:fork} and \cref{thm:abab-hardness} complete the complexity dichotomy for \lcol{$q$} in $H$-free ordered graphs for all $q \geq 4$.

\cref{thm:abab-hardness} would be nicely complemented by a polynomial-time algorithm for \col{$3$} in $\abab$-free ordered graphs, but we were unable to find it. Instead, we construct a quasipolynomial-time algorithm.

\begin{restatable}{theorem}{thmababalg}\label{thm:abab-algorithms}
    For every fixed $k \geq 0$, \lcol{3} in $n$-vertex \ababk-free ordered graphs can be solved in time $n^{\Oh(\log^4 n)}$.    
\end{restatable}

Let us emphasize that \cref{thm:abab-algorithms} combined with \cref{thm:abab-outerstring} implies that \lcol{$3$} can be solved in quasipolynomial time in outerstring graphs.
This result heavily extends the recent algorithm of E.S., Ganian, Lokshtanov, and Surianarayanan~\cite{DBLP:conf/sosa/SGLS26} for \lcol{$3$} in circle graphs, which are a subclass of outerstring graphs.
Intrestingly, the crux ideas behind both approaches are very different.

Finally, we proceed to the last gap in \cref{fig:summary}: \lcol{$3$} in \abbak-free ordered graphs. We already knew that for $q\geq 4$ the \lcol{$q$} problem is \NP-hard in this class \cite{DBLP:conf/stacs/PiecykR26}. Now we show that it is also the case for $q=3$.

\begin{restatable}{theorem}{thmabbahard}\label{thm:abba-hardness}
    \lcol{3} is \NP-hard on \abba-free ordered graphs.
\end{restatable}

Altogether, \cref{thm:fork,thm:abab-hardness,thm:abba-hardness} complete the complexity dichotomy for \lcol{$q$} in $H$-free ordered graphs for all $q \geq 3$ into (quasi)polynomial-time solvable and \NP-hard cases, as summarized in \cref{fig:summary}.

\subparagraph{Revisiting some \NP-hard cases.}
Next, we revisit two families of \NP-hard cases defined by forbidding a fixed (extension of) ordered two-edge matching, i.e., \abba and \abab.

First, we show that \lcol{3} in \abbak-free ordered graphs can be solved in subexponential time.

\begin{restatable}{theorem}{thmabbaalg}\label{thm:abba-algorithms}
    Let $k \geq 0$ be a fixed integer.
    \lcol{3} in $n$-vertex \abbak-free graphs can be solved in time $n^{\Oh(n^{2/3} \log n)}$.
\end{restatable}

It is somewhat unexpected, as in unordered $H$-free graphs, \lcol{$q$} is either (quasi)polynomial-time solvable or cannot be solved in subexponential time, unless the ETH fails.

Next, we consider \aabb.
It was shown in~\cite{DBLP:journals/siamdm/HajebiLS24} that already \lcol{3} is \NP-hard on \aabb-free ordered graphs.
We first show that the same holds for the non-list version of the problem.

\begin{restatable}{theorem}{thmaabbhard}\label{thm:aabb-hardness}
    \col{3} is \NP-hard on \aabb-free ordered graphs.
\end{restatable}

Then, we show two algorithms for the promise variant of the problem: if we are given a $3$-colorable \aabb-free ordered graph $G$, we can find a $5$-coloring of $G$ in polynomial time, or a $4$-coloring of $G$ in quasipolynomial time.

\begin{restatable}{theorem}{thmaabbalgo}\label{thm:aabb-algorithms}
    Given a \aabb-free ordered graph $G$, in polynomial (resp. quasipolynomial) time we can either decide that $\chi(G) > 3$, or find a proper 5-coloring (resp., 4-coloring) of $G$.
\end{restatable}

\section{Preliminaries}
\paragraph{Graph theory.}

For an ordered grap $G$ and a vertex $v$, by $N^+(v)$ we denote the set $N(v) \cap \{u \mid v \prec u\}$.

For two disjoint sets $A,B \subseteq V(G)$, by $E(A,B)$ we denote the set of edges with one endpoint in $A$ and the other in $B$.

\paragraph{List Coloring.}

Let $(G,L)$ and $(G',L')$ be two instances of \lcol{$3$}.
We say that $(G',L')$ is a \emph{subinstance} of $(G,L)$ if $G'$ is an induced subgraph of $G$ and for every $v \in V(G')$ it holds that $L'(v) \subseteq L(v)$.

We also recall the following classical result of Edwards~\cite{DBLP:journals/tcs/Edwards86}.

\begin{theorem}\label{thm:edwards}
    Every instance of \textsc{List Coloring} where each list is of size at most 2 can be solved in polynomial time. For yes-instances, a proper list coloring can be found in polynomial time as well.
\end{theorem}

\paragraph{Reduction rules and branching scheme.}
Let us briefly describe some common tools that will be used in our quasi\-polynomial-time algorithms.
Let $(G,L)$ be an instance of \lcol{$3$}.

Whenever possible, we will apply the following reduction rules to simplify the instance $(G,L)$.
\begin{enumerate}[(R1)]    
    \item If there is a vertex with an empty list, then report that $(G,L)$ is a no-instance.
    \item If there is a vertex $v$ with a singleton list, say $L(v) = \{c\}$, then we can remove $v$ from the graph and remove color $c$ from the lists of all neighbors of $v$.
\end{enumerate}
These rules clearly preserve the answer to the instance $(G,L)$ and can be applied in polynomial time.
Thus, we from now on we will always assume that each vertex has a list of size at least 2.

Now let us describe the idea behind the branching; it is based on ideas from~\cite{DBLP:journals/algorithmica/BonnetR19,DBLP:conf/sosa/PilipczukPR21}.

Let $G$ be a graph and let $\cF$ be a family of subsets of $V(G)$.
For a set $X \subseteq V(G)$, by $\cF|_X$ we denote the family $\{F \in \cF \mid F \subseteq X\}$.
For a vertex $v$, by $d_G(v,\cF)$ we denote the number of sets in $\cF$ that intersect $N_G[v]$.

In what follows, we think of the subgraphs in the families $\cF_1,\ldots,\cF_t$ as \emph{obstacles} that make the current instance complicated.
If we manage to reduce the instance by removing some vertices so that one of these families becomes empty, we obtain some significant progress towards a solution.

\begin{lemma}\label{lem:branching}
    Let $t,d \geq 1$ be fixed integers and $\epsilon \in (0,1)$ be a fixed real.
    Let $(G,L)$ be an $n$-vertex instance of \lcol{$3$} and let $\cF_1,\ldots,\cF_t$ be families of subsets of $V(G)$ such that, for every $i \in [t]$,
    each set in $\cF_i$ has at most $d$ vertices.

    Suppose that for every induced subgraph $G'$ of $G$ 
    there is $i \in [t]$ such that 
    either (i) $\cF_i|_{V(G')} = \emptyset$ 
    or (ii) there is a vertex $v \in \bigcup_{F \in \cF_i|_{V(G')}} F$ satisfying
    $d_{G'}(v,\cF_i|_{V(G')}) \geq \epsilon |\cF_i|_{V(G')}|$.
    Then in time $n^{\Oh(\log n)}$ we can either reject $(G,L)$ or enumerate a family $\cG$ of triples $(G',L',\phi')$, such that:
    \begin{enumerate}[(1)]
        \item for each $(G',L',\phi') \in \cG$:
        \begin{itemize}
            \item $(G',L')$ is a subinstance of $(G,L)$,
            \item $\phi' : V(G) \setminus V(G') \to [3]$ is a proper coloring of $G - V(G')$ that respects the lists $L$,
            and for every edge $uv \in E(G)$ with $u \in V(G')$ and $v \in V(G) \setminus V(G')$, it holds that $\phi'(v) \notin L'(u)$.   
        \end{itemize}  \label{it:subinstance}
        \item for each $(G',L',\phi') \in \cG$ there is $i \in [t]$ such that $\cF_i|_{V(G')} = \emptyset$, and \label{it:emptyfamily}
        \item $(G,L)$ is a yes-instance of \lcol{$3$} if and only if there is $(G',L',\phi') \in \cG$ such that $(G',L')$ is a yes-instance of \lcol{$3$}.\label{it:equivalent}
    \end{enumerate}
    In particular, for $(G',L',\phi')$ as in item~\ref{it:equivalent},
    any proper coloring of $G'$ that respects the lists $L'$
    can be combined with $\phi'$ to obtain a proper coloring of $G$ that respects the lists $L$.
\end{lemma}
\begin{proof}
    During the execution of the algorithm, we will apply reduction rules (R1) and (R2). Note that the application of (R2) corresponds to coloring a vertex $v$ with color $c$ and removing it from the current instance.
    These assigned colors will define the coloring $\phi'$ of $G - V(G')$ in the output. We will not explicitly store $\phi'$ during the execution of the algorithm, but we will be able to reconstruct it from the sequence of applied reduction rules (R2).

    Define $\cF = \bigcup_{i \in [t]} \cF_i$.
    For each $F \in \cF$, we enumerate all possible colorings of $G[F]$ that respect the lists $L$.
    If, for some $F$, there is no such coloring, then we can report that $(G,L)$ is a no-instance and terminate.
    Also, if for some $F$, some $v \in F$, and some $c \in L(v)$, there is no coloring of $G[F]$ that respects the lists $L$ and assigns color $c$ to $v$, then we can safely remove color $c$ from the list of $v$.

    We apply these steps, as well as the reduction rules, exhaustively. 
     As $|F| \leq d$ and $d$ is a constant, this can be done in polynomial time.If (R1) applies, then we report that $(G,L)$ is a no-instance and terminate.
       For simplicity, we keep calling the current instance $(G,L)$ and the obstacle families $\cF_1,\ldots,\cF_t$.

    For each $i \in [t]$, let $\cF_i^*$ be the family of all pairs $(F,\phi)$, where $F \in \cF_i$ and $\phi$ is a coloring of $G[F]$ that respects the lists $L$.
    Note that by the previous step, every $F \in \cF_i$ gives rise to at least one and at most $3^d$ elements of $\cF_i^*$.
    Consequently, $|\cF_i| \leq |\cF_i^*| \leq 3^d |\cF_i| = \Oh(|\cF_i|)$, and we can enumerate $\cF_i^*$ in polynomial time.

    If there is $\cF_i$ that is empty, then we can return $\cG = \{ (G,L,\phi') \}$, where $\phi'$ is the coloring defined by the applications of (R2), and terminate.
    So, suppose otherwise. By the assumption of the lemma applied to $G' = G$,
    there is $i \in [t]$ and a vertex $v$ such that $d_G(v,\cF_i) \geq \epsilon |\cF_i|$. We call $v$ the \emph{pivot}.

    Let $\cF_i(v) \subseteq \cF_i$ consist of those sets that intersect $N_G[v]$.
    From each of these sets pick a \emph{representative}: one vertex that is in $N_G[v]$.
    As each vertex has one of four possible lists,
    there is a family $\cF'_i(v) \subseteq \cF_i(v)$ such that all its representatives have the same list, say $A$, and $|\cF'_i(v)| \geq \frac{1}{4} |\cF_i(v)| \geq \frac{\epsilon}{4} |\cF_i|$.

    Now notice that there is a color $c \in A \cap L(v)$.
    We branch on whether $v$ is colored with $c$ or not.
    More precisely, we create two instances of \lcol{$3$}.
    In the first instance, we remove color $c$ from the list of $v$; we call this branch the \emph{failure branch}.
    In the second instance, we set the list of $v$ to $\{c\}$; we call this branch the \emph{success branch of type $i$}.
    In both instances, we exhaustively apply the reduction rules.

    Note that in case of a success branch, reduction rule (R2) removes $c$ from the lists of all neighbors of $v$ and also the vertex $v$ from the graph.
    So, every set $F \in \cF'_i(v)$ disappears from the resulting instance (if $v$ is its representative) or its representative loses a color from its list.
    In the latter case, at least one of the possible colorings of $F$ kept in $\cF_i^*$ is no longer available.
    Consequently, in the resulting instance, the size of the updated set $\cF_i^*$ is at most $(1-\frac{\epsilon}{4 \cdot 3^d}) |\cF_i^*|$, i.e., it shrinks by a constant factor.
    
    We apply the same procedure recursively to each of the two instances, discarding a branch when (R1) rejects it and terminating a surviving branch when one of the families $\cF_i$ becomes empty.
    In any other case, we are guaranteed to find an appropriate pivot vertex $v$.
    We let $\cG$ consist of the surviving leaves of the recursion tree,
    together with the colorings of removed vertices defined by the applications of (R2).

    Clearly, items \ref{it:subinstance}, \ref{it:emptyfamily}, and \ref{it:equivalent} in the lemma statement are satisfied. Let us argue that the running time and the size of the family $\cG$ are as claimed.
    Consider the recursion tree $\mathsf{T}$ of the algorithm.
    As every node is processed in polynomial time, it is enough to bound the number of leaves of $\mathsf{T}$.

    Each edge of $\mathsf{T}$ corresponds either to a failure branch, or to one of $t$ types of success branches.
    Edges corresponding to failure branches form upward paths, each with at most $3n$ vertices, as in each failure branch we remove a color from a vertex in the graph.

    Contract every edge corresponding to a failure branch, and let $\mathsf{T}'$ be the resulting tree.
    Each node of $\mathsf{T}'$ has degree $\Oh(n)$.
    In each root-to-leaf path of $\mathsf{T}'$, the number of success branches of type $i$ is at most $\Oh(\log |\cF_i|)=\Oh(\log n^d)=\Oh(\log n)$ as each such branch reduces the total size of $\cF_i^*$ by a constant factor.
    As the number $t$ of types is constant, $\mathsf{T}'$ has at most $n^{\Oh(\log n)}$ leaves.
    The number of leaves in $\mathsf{T}$ is at most $\Oh(n)$ times the number of leaves in $\mathsf{T}'$, i.e., $n^{\Oh(\log n)}$.
    This completes the proof of the lemma.
\end{proof}

\section[Excluding small extensions of a forward fork]{Excluding \kjj}
In this section we prove \cref{thm:fork}. 
Before we proceed to the proof, let us recall some well-known properties of treewidth. Here, a vertex is called \emph{simplicial} if its neighborhood induces a clique.

\begin{proposition}\label{prop:tw}
    Let $G$ be an unordered graph.
    \begin{enumerate}
        \item For every $S \subseteq V(G)$, it holds that $\tw(G) \leq \tw(G - S) + |S|$.
        \item If $v$ is a simplicial vertex of $G$, then $\tw(G) \leq \max (|N[v]|-1, \tw(G - v))$.
        \item For any edge $e \in E(G)$, it holds that $\tw(G-e) \leq \tw(G)$.
    \end{enumerate}
\end{proposition}
\begin{proof}
    To see the first item, consider a tree decomposition of $G-S$ with width $\tw(G-S)$ and add every vertex from $S$ to every bag. This way we obtain a tree decomposition of $G$ with width $\tw(G-S) + |S|$.

    To see the second item, consider a tree decomposition of $G-v$ with width $\tw(G-v)$. Since $N(v)$ is a clique, the properties of tree decompositions imply that there is a node whose bag contains this set. Add a new node, adjacent to the node with the bag containing $N(v)$, and assign to this new node the bag $N[v]$. This way we obtain a tree decomposition of $G$ with width $\max (|N[v]|-1, \tw(G-v))$.

    The last property is obvious, since any tree decomposition of $G$ is also a tree decomposition of $G-e$.
\end{proof}

We will also use the following observation~\cite{DBLP:journals/siamdm/HajebiLS24}.

\begin{lemma}\label{lem:chordal}
    Every \jj-free graph is chordal.
\end{lemma}

The following lemma is the crucial combinatorial observation used in our algorithm.

\begin{lemma}\label{lem:fork-tw}
    Let $k,q,r$ be positive integers and let $G$ be a \kjj-free ordered graph with clique number at most $q$.
    Let $X_1,\ldots,X_r$ be (possibly empty and not necessarily disjoint) subsets of $V(G)$ such that $\bigcup_{i \in [r]} X_i = V(G)$ and for each $i \in [r]$, the graph $G[X_i]$ is \jj-free.
    Let $F$ be the graph with vertex set $V(G)$ and edge set $\bigcup_{i \in [r]} E(G[X_i])$.
    Then $\tw(F) = \Oh(kqr^3)$.
    Furthermore, a tree decomposition of $F$ of width $\Oh(kqr^3)$ can be found in polynomial time.
\end{lemma}
\begin{proof}
    Define 
    \begin{align*}
        w_1 & = q-1,\\
        w_p & = w_{p-1} + kq \binom{r}{2} + 2k(q-1) \text{ for } p \geq 2.
    \end{align*}
Note that $w_r = \Oh(kqr^3)$ and that $q-1 = w_1 \leq w_2 \leq \ldots \leq w_r$.
The proof is by induction on $r$.
For $r=1$ we have $F=G$. As $G$ is \jj-free, by \cref{lem:chordal} it is chordal, and thus its treewidth is equal to the clique number minus 1, which is at most $q -1 = w_1$.
So, suppose that $r \geq 2$ and the statement holds for $r-1$.
Define 
\[
 M = \{ v \in V(G) \mid v \text{ belongs to at least two sets } X_i\}
\]
Consider two cases.

\noindent\textbf{Case 1: $|M| < kq \binom{r}{2}$.}
Every vertex of $F-M$ belongs to exactly one set $X_i$, and thus $F-M$ is the disjoint union of the graphs $G[X_i]-M$ for $i \in [r]$. As each of these graphs is chordal, so is $F-M$. As every clique of $F$ is a clique of $G$, the clique number of $F-M$ is at most $q$, and thus $\tw(F-M) \leq q-1$.
By \cref{prop:tw}, we have $\tw(F) \leq \tw(F-M) + |M| \leq q - 1 + kq \binom{r}{2} \leq w_r$.

\medskip
\noindent\textbf{Case 2: $|M| \geq kq \binom{r}{2}$.}
Let $B$ be the first $kq \binom{r}{2}$ vertices of $M$ with respect to $\prec$, and let $b$ be the last vertex from $B$.
To each vertex $v \in B$, we assign a pair $\{i,j\}$ of distinct indices in $[r]$ such that $v \in X_i \cap X_j$ (if there are multiple such pairs, we choose one arbitrarily).
As the number of possible pairs is $\binom{r}{2}$, by the pigeonhole principle there is a pair $\{i,j\}$ such that at least $kq$ vertices of $B$ are assigned to this pair.
By symmetry, we can assume that $i=r-1$ and $j=r$.
Let $T \subseteq B \cap X_{r-1} \cap X_r$ be the set of $kq$ vertices assigned to the pair $\{r-1,r\}$.

\begin{claim}
    There is an independent set $A \subseteq T$ of size $k$.
\end{claim}
\begin{claimproof}
    We pick this set greedily, processing the vertices of $T$ from left to right: we take the smallest available vertex and discard it together with all its forward neighbors.

    As $T \subseteq X_r$, the graph $G[T]$ is \jj-free and thus,
    restricted to this graph, forward neighbors of every vertex $v$, together with this vertex, form a clique.
    Consequently, at each step we discard at most $q$ vertices.
    Hence, we can pick $k$ vertices before we run out of vertices in $T$, as desired.
\end{claimproof}

Define
\begin{align*}
    D & = \{ u \in X_{r-1} \cup X_r \mid b \prec u \text{ and } u \text{ has a neighbor in } A\},\\
    Z & = B \cup D.
\end{align*}
For every $a \in A$, its forward neighbors in $X_{r-1}$ (resp. $X_{r}$) together with $a$ form a clique, and thus $a$ has at most $2(q-1)$ forward neighbors in $X_{r-1} \cup X_r$.
As $A \subseteq B$ and every vertex of $D$ succeeds $b$, every neighbor of $a$ in $D$ is a forward neighbor of $a$.
Consequently, as $|A| = k$, we have
\begin{align*}
    |D| & \leq 2k(q-1),\\
    |Z| & \leq kq \binom{r}{2} + 2k(q-1).
\end{align*}
We partition $V(G) \setminus Z$ into \emph{prefix} $P$ and \emph{suffix} $S$ as follows:
\begin{align*}
    P & = \{ v \in V(G) \setminus B \mid v \prec b\},\\
    S & = \{ v \in V(G) \setminus D \mid v \succ b\}.
\end{align*}
At this point, we partitioned the vertex set of $F$ into three sets $P$, $Z$, and $S$. We will deal with them one by one, starting with $S$.

\begin{claim}
    The graph $G[(X_{r-1} \cup X_r) \cap S]$ is \jj-free.
\end{claim}
\begin{claimproof}
    Suppose for a contradiction that there is a set $Q$ of three vertices that induces a copy of \jj in $G[(X_{r-1} \cup X_r) \cap S]$.
    Note that $A \prec Q$ and there are no edges between these sets, as $D \cap S = \emptyset$.
    Consequently, $A \cup Q$ induces a copy of \kjj in $G$, which is a contradiction.
\end{claimproof}

We apply the induction hypothesis to the graph $G[S]$ with $r-1$ sets $X_1 \cap S, \ldots, X_{r-2} \cap S, (X_{r-1} \cup X_r) \cap S$; the corresponding graph is the graph $F'$ obtained from $F[S]$ by adding all edges of $G$ between $X_{r-1} \cap S$ and $X_r \cap S$.
As treewidth does not increase when we remove edges, we have $\tw(F[S]) \leq \tw(F') \leq w_{r-1}$.

Now let us focus on $P$. As vertices of $P$ precede $b$ and do not belong to $B$, every $v \in P$ belongs to exactly one set $X_i$ for $i \in [r]$. Consequently, all neighbors of $v$ in $F$ belong to $X_i$ and its forward neighbors form a clique in $F$; in particular, there are at most $q-1$ of them.

Let $p_1 \prec p_2 \prec \ldots \prec p_m$ be the vertices of $P$ and consider the graph $F - Z = F[P \cup S]$.
For every $j \in [m]$, all neighbors of $p_j$ in the graph $F[\{p_j,\ldots,p_m\} \cup S]$ are forward neighbors of $p_j$: indeed, every vertex of $S$ succeeds $b$ and thus succeeds $p_j$, while the vertices $p_1,\ldots,p_{j-1}$ have already been removed.
Hence $p_j$ is a simplicial vertex of that graph and its closed neighborhood is of size at most $q$.
Applying the second item of \cref{prop:tw} to $p_1,\ldots,p_m$, in this order, we obtain
\[
\tw(F[P \cup S]) \leq \max(q-1, \tw(F[S])) \leq \max(q-1,w_{r-1}) = w_{r-1}.
\]
Finally, as $|Z| \leq kq \binom{r}{2} + 2k(q-1)$, the first item of \cref{prop:tw} gives
\[
\tw(F) \leq \tw(F - Z) + |Z| \leq w_{r-1} + kq \binom{r}{2} + 2k(q-1) = w_r.
\]
The proof can be easily turned into a polynomial-time algorithm: all steps can be performed in polynomial time and the recursion consists of a single recursive call, applied to an instance with one set less.
This completes the proof of the lemma.
\end{proof}

Finally, we are ready to prove \cref{thm:fork}.

\thmfork*
\begin{proof}    
    \setcounter{claim}{2}    
    Let $(G,L)$ be an instance of \lcol{$q$}, where $G$ is a \kjj-free ordered graph.
    For simplicity of presentation, let us add to $G$ a set of $qk$ new isolated vertices, placed at the end of the ordering, and set their lists to singletons so that, for each $i \in [q]$,
    exactly $k$ new vertices have list $\{i\}$.
    This results in an equivalent instance of \lcol{$q$}, as isolated vertices with nonempty lists can always be colored.
    Furthermore, \kjj-freeness is preserved: a new vertex cannot play the role of any of the three vertices of the copy of \jj, as each of them has a neighbor in the copy, and it cannot play the role of one of the $k$ initial vertices either, as it succeeds every other vertex.
    As both $k$ and $q$ are fixed, we can focus on solving the modified instance in polynomial time. 
    For simplicity, let us still denote it by $(G,L)$.
    The purpose of the modification was to ensure that in every proper $q$-coloring respecting the lists, each color is used at least $k$ times.

    First, we exhaustively check if $G$ has a clique of size $q+1$. 
    In case it does, we conclude that $(G,L)$ is a no-instance and immediately reject. So, from now on we can assume that the clique number of $G$ is at most $q$.

    Exhaustively enumerate all tuples $\cY = (Y_1,\ldots,Y_q)$ of pairwise disjoint and independent subsets of $V(G)$, each of size exactly $k$.
    As $k$ and $q$ are fixed, there are $\Oh(n^{kq})$ such tuples and they can be enumerated in polynomial time.
    The intended meaning is that $Y_i$ consists of the first $k$ vertices in color $i$ in some fixed solution; recall that, by the modification above, each color is used at least $k$ times.
    Clearly, if a solution exists, there is a branch of the enumeration that corresponds to it.
    Thus, if any of the branches leads to a solution, we can return it as a solution to the original instance. On the other hand, if all branches fail, we can conclude that the original instance is a no-instance.

    Fix one tuple $\cY = (Y_1,\ldots,Y_q)$ and, for each $i \in [q]$, let $y_i$ be the last vertex from $Y_i$.
    We modify the lists of the vertices of $G$. For each $i \in [q]$ and each vertex $v \in V(G)$, we do the following:
    \begin{enumerate}
        \item If $v \in Y_i$, we remove from its list all colors except for $i$; note that the list becomes empty if $i \notin L(v)$.
        \item If $v \notin Y_i$ and $v \prec y_i$, we remove color $i$ from the list of $v$.
        \item If $v$ has a neighbor in $Y_i$, we remove color $i$ from the list of $v$.
    \end{enumerate}
    Let $L'$ denote the resulting lists. Clearly, they are consistent with the intended meaning of the sets $Y_i$.
    If any vertex ends up with an empty list, we can immediately reject the current branch. Otherwise, we proceed to the next step.

    Let $G' = G - \bigcup_{i \in [q]} Y_i$. Note that solving the current branch is equivalent to solving $(G',L')$ as an instance of \lcol{$q$}.
    Indeed, the colors of the vertices in $\bigcup_{i \in [q]} Y_i$ are fixed, and the lists of the remaining vertices are consistent with this coloring.

    As $G'$ is an induced subgraph of $G$, it is also \kjj-free and has clique number at most $q$.    
    For each $i \in [q]$, let $X_i = \{v \in V(G') \mid i \in L'(v) \}$. Note that the sets $X_1,\ldots,X_q$ cover $V(G')$ as no vertex has empty list.

    \begin{claim}
        For each $i \in [q]$, the graph $G'[X_i]$ is \jj-free.
    \end{claim}
    \begin{claimproof}
        For a contradiction, suppose that there is a set $Q$ of three vertices that induces a copy of \jj in $G'[X_i]$.
        As $Q \subseteq X_i$, every vertex of $Q$ has $i$ in its modified list. Thus, by the third rule, there are no edges between $Y_i$ and $Q$ and, by the second rule, $y_i \prec v$ for every $v \in Q$, i.e., $Y_i \prec Q$.
        As $Y_i$ is independent and of size $k$, we obtain an induced copy of \kjj in $G$, a contradiction.
    \end{claimproof}

    Let $F$ be the graph with vertex set $V(G')$ and edge set $\bigcup_{i \in [q]} E(G'[X_i])$ and note that $(F,L')$ is a yes-instance of \lcol{$q$} if and only if $(G',L')$ is a yes-instance of \lcol{$q$}. Indeed, the endpoints of every removed edge have disjoint lists in $L'$, so such an edge can never be monochromatic and thus it is irrelevant for the coloring problem.
    We remark that in general removing edges is likely to spoil \kjj-freeness, but we are not going to use it anymore.
    Instead, we invoke \cref{lem:fork-tw}, applied with $r = q$, to obtain a tree decomposition of $F$ of width $\Oh(kq^4)$, which is a constant as both $k$ and $q$ are fixed.

    Consequently, $(F,L')$ can be solved in polynomial time by a straightforward dynamic programming on a tree decomposition~\cite{DBLP:books/sp/CyganFKLMPPS15}.
    Summing up, the algorithm considers $\Oh(n^{kq})$ branches and processes each of them in polynomial time, so its total running time is $n^{\Oh(kq)}$. This completes the proof.
\end{proof}

\section[Excluding two crossing edges or their extension]{Excluding \abab or its extensions}

\subsection{Algorithms}
In this section we show the following result.

\thmababalg*

We prove \cref{thm:abab-algorithms} in a few steps.

\paragraph{Destroying crossing pairs.}
Let $G$ be an ordered graph and let $X,Y$ be two disjoint subsets of $E(G)$.
We say that $(X,Y)$ is a \emph{crossing pair} if there are subsets $A,B,C,D$ of $V(G)$ such that $A \prec B \prec C \prec D$, every edge in $X$ has one endpoint in $A$ and the other in $C$, and every edge in $Y$ has one endpoint in $B$ and the other in $D$.

\begin{lemma}\label{lem:crossingXY}
    Let $k \geq 0$ be a fixed constant.
    Let $(G,L)$ be an $n$-vertex instance of \lcol{$3$} where $G$ is \ababk-free.
    Let $(X,Y)$ be a crossing pair in $G$.
    In time $n^{\Oh(\log n)}$ we can either reject $(G,L)$ or enumerate a family $\cF$ of at most $n^{\Oh(\log n)}$ pairs $(G',L')$ such that:
    \begin{itemize}
        \item every $(G',L') \in\cF$ is a subinstance of $(G,L)$,
        \item for every $(G',L') \in\cF$, either $X \cap E(G') = \emptyset$ or $Y \cap E(G') = \emptyset$,
        \item $(G,L)$ is a yes-instance of \lcol{$3$} if and only if there exists $(G',L') \in \cF$ such that $(G',L')$ is a yes-instance of \lcol{$3$}.
    \end{itemize}
\end{lemma}
\begin{proof}
    If $X = \emptyset$ or $Y = \emptyset$, then we can return the instance $(G,L)$ as the only element of $\cF$.
    Thus, suppose that $X \neq \emptyset$ and $Y \neq \emptyset$.
    Let $A \prec B \prec C \prec D$ be the sets of vertices witnessing that $(X,Y)$ is a crossing pair.
    Define
    \begin{align*}
        p = & \ 2k+1\\
        q = & \ 5k+1\\
        \epsilon = & \ \frac{1}{14k+4}.
    \end{align*}
    Recall that, for $v \in V(G)$ and $Z \in \{X,Y\}$, we have
    \[
        d(v,Z) = | \{ e \in Z \mid N[v] \cap e \neq \emptyset \} |.
    \]

    \begin{claim}\label{clm:heavyvertex}
        There exists a vertex $v$ incident with an edge from $X \cup Y$ such that $d(v,X) \geq \epsilon |X|$ or $d(v,Y) \geq \epsilon |Y|$.
    \end{claim}
    \begin{claimproof}
        For contradiction, suppose that for every vertex $v$ incident with an edge from $X \cup Y$ it holds that $d(v,X) < \epsilon |X|$ and $d(v,Y) < \epsilon |Y|$.
        
        We greedily pick a set $M_X \subseteq X$ of $p$ pairwise disjoint and nonadjacent edges.
        We can do it, as every edge in $X$ is adjacent to fewer than $2\epsilon |X|$ edges from $X$ and $2p\epsilon |X| \leq |X|$.
        Let $Y'$ be the set of edges in $Y$ that are not adjacent to any edge in $M_X$.
        Note that $|Y'| \geq |Y| - 2p\epsilon |Y|$.

     Then, greedily pick a set $M_Y\subseteq Y'$ of $q$ pairwise disjoint and nonadjacent edges. Every edge of $Y'$ is adjacent to fewer than $2\epsilon |Y|$ edges of $Y$.
     Moreover, no edge of $Y'$ is adjacent to an edge of $M_X$, by the definition of $Y'$.
     Therefore, each selected edge excludes fewer than $2\epsilon |Y|$ further edges of $Y'$, and the greedy construction succeeds because
     \[
        2q\epsilon |Y| = (1-2p\epsilon)|Y| \leq |Y'|.
    \]

    Consequently, the edges from $M_X \cup M_Y$ form an induced matching.
    Let $a$ be the $(k+1)$-th vertex of $V(M_X) \cap A$.
    Note that there are $k$ vertices in $V(M_X) \cap A$ preceding $a$, and another $k$ vertices in $V(M_X) \cap A$ following $a$, and all these vertices form an independent set.
    Let $c \in C$ be the vertex such that $ac \in M_X$.

    Now, pick an edge $bd \in M_Y$, where $b \in B$ and $d \in D$, so that:
    \begin{itemize}
            \item there are at least $k$ vertices of $V(M_Y) \cap B$ following $b$, 
            \item there are at least $2k$ vertices of $V(M_Y) \cap D$ preceding $d$, and
            \item there are at least $2k$ vertices of $V(M_Y) \cap D$ following $d$.
    \end{itemize}
    Such an edge exists as the conditions above exclude at most $5k$ edges from $M_Y$ while $|M_Y|=q = 5k+1$.
    The vertices $a,b,c,d$ induce \abab in $G$.
    Let $I_1$ be the set of $k$ vertices of $V(M_X) \cap A$ preceding $a$,
    $I_2$ be the set of $k$ vertices of $V(M_X) \cap A$ following $a$,
    and let $I_3$ be a set of $k$ vertices of $V(M_Y) \cap B$ following $b$.
    As there are at least $2k$ vertices of $V(M_Y) \cap D$ preceding $d$, we can pick $k$ of them that are not adjacent to $I_3$; call this set $I_4$.
    Similarly, we pick a set $I_5$ of $k$ vertices of $V(M_Y) \cap D$ following $d$ and nonadjacent to $I_3$.
    The set $I_1 \cup I_2 \cup I_3 \cup I_4 \cup I_5$ is an independent set, and together with $a,b,c,d$ it induces \ababk, a contradiction.
    \end{claimproof}

    Note that the vertex $v$ from \cref{clm:heavyvertex} can be found in polynomial time.

    For every induced subgraph $G'$ of $G$, if neither $X \cap E(G')$ nor $Y \cap E(G')$ is empty, then \cref{clm:heavyvertex} supplies a vertex satisfying the second outcome in the assumption of \cref{lem:branching}.
    Thus, the assumptions of \cref{lem:branching} are satisfied for $t=d=2$, $\epsilon$, and the obstacle families $X,Y$.
    We apply \cref{lem:branching}. If it rejects $(G,L)$, then we do the same.
    Otherwise, we return the pairs $(G',L')$ obtained from its output triples $(G',L',\phi')$ by forgetting $\phi'$.
\end{proof}

\paragraph{Chains.}
Let $(G,L)$ be an instance of \lcol{$3$} where $G$ is an ordered graph.
A \emph{chain} $\cQ$ over $G$ is a partition of $V(G)$ into $\ell \geq 3$ consecutive segments $Q_0,\ldots,Q_{\ell-1}$ called \emph{links}, such that every edge of $G$ is contained in a single link or joins two consecutive links, where $Q_0$ and $Q_{\ell-1}$ are considered consecutive.
All operations on indices of links are taken modulo $\ell$. 

For two chains $\cQ = (Q_0,\ldots,Q_{\ell-1})$ and $\cR = (R_0,\ldots,R_{\ell})$,
we say that $\cR$ is a \emph{refinement of $\cQ$ at $i$} if:
\begin{itemize}
    \item $R_j \subseteq Q_j$ for $0 \leq j < i$,
    \item $R_i \cup R_{i+1} \subseteq Q_i$,
    \item $R_j \subseteq Q_{j-1}$ for $i+1 < j \leq \ell$.
\end{itemize}    
In other words, $\cR$ is obtained from $\cQ$ by splitting the $i$-th link into two consecutive links and possibly removing some vertices.
A chain is a \emph{refinement} of another chain if it can be obtained by a sequence of refinements at links.

Note that every ordered graph can be seen as a chain of length 3 by splitting $V(G)$ arbitrarily into three consecutive segments.
Thus, throughout this section, we will (almost) always assume that the input instance of \lcol{$3$} is given along with its chain structure $\cQ$,
and will refer to the input instance as $(\cQ,L)$.
In particular, when it does not lead to confusion, we will identify a chain with its underlying ordered graph.

\begin{lemma}\label{lem:refinechain}
    Let $k \geq 0$ be a fixed constant.
    Let $(\cQ,L)$ be an $n$-vertex instance of \lcol{$3$}
    where $\cQ=(Q_0,\ldots,Q_{\ell-1})$ is \ababk-free.
    Let $i \in \{0,\ldots,\ell-1\}$.

    In time $n^{\Oh(\log^2 n)}$ we can either reject $(\cQ,L)$ or enumerate a family $\cF$ of at most $n^{\Oh(\log^2 n)}$ pairs $(\cR,\widehat{L})$ such that:
    \begin{itemize}
        \item every $(\cR,\widehat{L}) \in \cF$ is a subinstance of $(\cQ,L)$ where $\cR$ is a refinement of $\cQ$ at $i$,
        \item $(\cQ,L)$ is a yes-instance of \lcol{$3$} if and only if there exists $(\cR,\widehat{L}) \in \cF$ such that $(\cR,\widehat{L})$ is a yes-instance of \lcol{$3$}.
    \end{itemize}
\end{lemma}
\begin{proof}
    Throughout the proof, we will always assume that the current instance is reduced with respect to reduction rules (R1) and (R2).

    The aim is to split the $i$-th link into two consecutive links.
    We will do it by recursively cleaning the edges between $Q_i$ and its neighboring links.
    The state of the recursion consists of a current subinstance $(\cQ',L')$, where $\cQ'=(Q_0',\ldots,Q'_{\ell-1})$ and, for each $j \in \{0,\ldots,\ell-1\}$, we have $Q'_j \subseteq Q_j$, and a partition of $Q'_i$ into three consecutive segments $A,B,C$ such that $A \prec B \prec C$ and
    \[
        E(A,Q'_{i+1}) = \emptyset \qquad\text{and}\qquad E(C,Q'_{i-1}) = \emptyset.
    \]    
    We initially have $\cQ'=\cQ$, $L'=L$, $A=C=\emptyset$, and $B=Q_i$.

    If $B=\emptyset$, then we output the subinstance with chain
    \[
        (Q'_0,\ldots,Q'_{i-1},A,C,Q'_{i+1},\ldots,Q'_{\ell-1})
    \]
    and lists $L'$. This is a refinement of $\cQ$ at link $i$.
    Indeed, the invariant excludes precisely the two kinds of edges that would join nonconsecutive links after splitting $Q'_i$ into $A$ and $C$.
    Thus, our goal is to reduce the size of $B$ until it becomes empty.

    Suppose next that $B=\{v\}$.
    Since (R1) cannot be applied, we have $L'(v) \neq \emptyset$.    
    For every $c \in L'(v)$, we create a corresponding instance by setting the list of $v$ to $\{c\}$ and applying reduction rules exhaustively. Note that (R2) will remove vertex $v$ from the current instance.
    This way, we create at most three instances, and in each of them the corresponding set $B$ is empty.
    Thus, they can be immediately included in the returned family $\cF$.

    It remains to consider a state with $m=|B| \geq 2$.
    Let $\overleftarrow{B}$ and $\overrightarrow{B}$ be the first $\lfloor m/2 \rfloor$ and the last $\lceil m/2 \rceil$ vertices of $B$, respectively, and define
    \[
        X = E(\overleftarrow{B},Q'_{i+1}) \qquad\text{and}\qquad Y = E(\overrightarrow{B},Q'_{i-1}).
    \]
    We observe that $(X,Y)$ or $(Y,X)$ is a crossing pair and thus, we can apply \cref{lem:crossingXY}.
    If it rejects the current instance, then we discard the state.
    Otherwise, let $(G'',L'')$ be one of the returned subinstances
    and let $\cQ''=(Q''_0,\ldots,Q''_{\ell-1})$ be a chain over $G''$ obtained from $\cQ'$ by removing all vertices that are not in $G''$.
    Define $A'' = A \cap V(G'')$ and analogously $\overleftarrow{B}'',\overrightarrow{B}''$, and $C''$.
    
    If $X \cap E(G'')=\emptyset$, we continue with the state
    \[
        (\cQ'',L'') \text{ and } (A'' \cup \overleftarrow{B}'',\overrightarrow{B}'', C'') \text{ as the partition of } Q''_i.
    \]
    Otherwise, $Y \cap E(G'')=\emptyset$, and we continue with the state
    \[
        (\cQ'',L'') \text{ and } (A'', \overleftarrow{B}'',\overrightarrow{B}'' \cup  C'') \text{ as the partition of } Q''_i.
    \]
    In both cases, the structure of the partition is preserved and the new middle segment has size at most $\lceil m/2 \rceil$.
    We continue this process until the middle segment becomes empty, at which point we output the corresponding subinstance.

    Every replacement above preserves the property that the original instance is a yes-instance if and only if some current state is a yes-instance.
    In a root-to-leaf path, the middle segment is reduced at most $\Oh(\log n)$ times before its size becomes at most 1. By \cref{lem:crossingXY}, each nonterminal state produces at most $n^{\Oh(\log n)}$ states. Hence, the recursion tree has $n^{\Oh(\log^2 n)}$ nodes, and the same bound holds for its total running time and for the number of output instances.
\end{proof}

In the next lemma we will start with a three-link chain $\cQ = (A,B,C)$ and repeatedly refine it.
Note that if $\cQ'$ is a refinement of $\cQ$, then every link of $\cQ'$ originates at one of the original links $A,B,C$.
This way we define \emph{types} of links of $\cQ'$: a link is of type $\lambda \in \{A,B,C\}$ if it originates from the original link $\lambda$ of $\cQ$.

\begin{lemma}\label{lem:breakchain}
    Let $k \geq 0$ be a fixed constant.
    Let $(\cQ,L)$ be an $n$-vertex instance of \lcol{$3$}
    where $\cQ=(A,B,C)$ is \ababk-free.
    
    In time $n^{\Oh(\log^3 n)}$ we can either reject $(\cQ,L)$ or enumerate a family $\cF$ of at most $n^{\Oh(\log^3 n)}$ pairs $(\cR,\widehat{L})$ such that:
    \begin{itemize}
        \item every $(\cR,\widehat{L}) \in \cF$ is a subinstance of $(\cQ,L)$ where $\cR$ is a refinement of $\cQ$,
        \item for every $(\cR,\widehat{L}) \in \cF$, the chain $\cR$ has at least one empty link of each type $A,B,C$,
        \item $(\cQ,L)$ is a yes-instance of \lcol{$3$} if and only if there exists $(\cR,\widehat{L}) \in \cF$ such that $(\cR,\widehat{L})$ is a yes-instance of \lcol{$3$}.
    \end{itemize}
\end{lemma}
\begin{proof}
    We will use \cref{lem:refinechain} to recursively refine $\cQ$.
    The state of the recursion is thus a subinstance $(\cQ',L')$ where $\cQ'$ is some refinement of $\cQ$; initially $(\cQ',L') = (\cQ,L)$.    
    
    Consider one such state $(\cQ',L')$.
    If $\cQ'$ has at least one empty link of each type, then we include it in the family $\cF$ and terminate the recursion.
    So suppose that there is $\lambda \in \{A,B,C\}$ such that $\cQ'$ has no empty link of type $\lambda$.
    We apply \cref{lem:refinechain} to the smallest link of type $\lambda$ in $\cQ'$.
    Note that at each level of recursion, the size of a smallest link of type $\lambda$ halves,
    so there are at most $\Oh(\log n)$ levels of recursion corresponding to type $\lambda$.
    Since there are three types, the depth of the recursion is $\Oh(\log n)$.
    As \cref{lem:refinechain} branches into at most $n^{\Oh(\log^2 n)}$ subinstances,
    the total number of leaves of the recursion tree is at most $n^{\Oh(\log^3 n)}$.
\end{proof}

Finally, we are ready to show \cref{thm:abab-algorithms}.

\thmababalg*
\begin{proof}
    Let $(G,L)$ be an $n$-vertex instance of \lcol{$3$}, where $G$ is \ababk-free.
    We can assume that $n \geq 3$, as otherwise we can solve the instance by brute force.

    We set up the initial chain over $G$ as $\cQ = (A,B,C)$,
    where $A$ consists of the first $\lfloor n/3 \rfloor$ vertices of $G$,
    $B$ consists of the next $\lfloor n/3 \rfloor$ vertices,
    and $C$ consists of the remaining vertices.
    
    We apply \cref{lem:breakchain} to $(\cQ,L)$.
    If it rejects the instance, then we do the same.
    Otherwise, consider the family $\cF$ of subinstances $(\cQ',L')$ returned by \cref{lem:breakchain}.
    
    Fix one such subinstance $(\cQ',L')$ and denote its links by $(A_1,\ldots,A_a,B_1,\ldots,B_b,C_1,\ldots,C_c)$ for some $a,b,c \geq 1$, where each $A_i$ is of type $A$ etc.

    Let $i_A$ be the smallest integer such that $A_{i_A} = \emptyset$; it exists by the guarantee of \cref{lem:breakchain}. In the same way we define $i_B$ and $i_C$.

    We partition $\cQ'$ into three chains $\cQ_1,\cQ_2,\cQ_3$ defined as follows:
    \begin{align*}
        \cQ_1 = & \ (A_1,\ldots,A_{i_A-1},C_{i_C+1},\ldots,C_c),\\
        \cQ_2 = & \ (A_{i_A+1},\ldots,A_a,B_1,\ldots,B_{i_B-1}),\\
        \cQ_3 = & \ (B_{i_B+1},\ldots,B_b,C_1,\ldots,C_{i_C-1}).
    \end{align*}    
    These chains are vertex-disjoint and cover all vertices of $\cQ'$.
    Furthermore, there are no edges between two different chains as edges exist only between consecutive links and we split $\cQ'$ at the empty links $A_{i_A},B_{i_B},C_{i_C}$.    
    Consequently, we can call the algorithm recursively on each of these three chains with $L'$ restricted to the vertices of each chain, and return a yes-instance if and only if all three recursive calls return a yes-instance.    

    We note that each of these three subinstances has at most $\lceil 2n/3 \rceil$ vertices, as each of them avoids one of the original links $A,B,C$.
    Consequently, the depth of recursion is $\Oh(\log n)$.
    As the family returned by \cref{lem:breakchain} is of size $n^{\Oh(\log^3 n)}$, we conclude that the total number of leaves of the recursion tree and the overall running time is at most $n^{\Oh(\log^4 n)}$.
  \end{proof}

\subsection{Connection to outerstring graphs and its consequences}
Let $D$ be a closed disk within some plane and let $\partial D$ denote its boundary.
An outerstring graph $G=(V,E)$ is the intersection graph of some family of curves $\{s(v)\}_{v\in V}$, contained within $D$, such that each $s(v)$ intersects $\partial D$ precisely once, at point $p(v)$ called its \emph{anchor}.

\thmababouterstring*
\begin{proof}
    Let $G$ be any outerstring graph. 
    We order the vertices of $G$ according to the order of their anchors appearing on $\partial D$, starting from an arbitrary point and going clockwise.
    This can clearly be done in time polynomial in the number of vertices.
    
    We claim that the resulting ordered graph is \abab-free.
    Suppose to the contrary that there is a copy of \abab induced by some vertices $a\prec b\prec c \prec d$.
    Note that the anchors $p(a),p(b),p(c),p(d)$ appear in this clockwise order on $\partial D$.
    As $ac$ is an edge of $G$, the curves $s(a)$ and $s(c)$ intersect at some point $p$.
    Let $s$ be the curve obtained by following $s(a)$ from $p(a)$ up to $p$ and then following $s(c)$ from $p$ to $p(c)$; note that $s$ is contained in $D$ and intersects $\partial D$ exactly at $p(a)$ and $p(c)$.
    Let $s'$ be the closed loop obtained from $s$ by following one of the two arcs of $\partial D$ from $p(c)$ back to $p(a)$.
    By the Jordan curve theorem, $s'$ divides the plane into two regions.
    Since the remaining part of $s'$ lies on $\partial D$, the curve $s$ divides $D$ into two regions, each containing one of the two arcs of $\partial D$ with endpoints $p(a)$ and $p(c)$.
    Thus $p(b)$ and $p(d)$ lie on different such arcs, and hence in different components of $D \setminus s$.
    As $bd$ is an edge of $G$, the curves $s(b)$ and $s(d)$ intersect, so their union contains a curve joining $p(b)$ and $p(d)$.
    This curve is contained in $D$, so it must cross $s$, and thus $s(a)$ or $s(c)$.
    Consequently, $b$ or $d$ is adjacent to $a$ or $c$, so the copy of \abab is not induced, a contradiction.
\end{proof}

Equipped with \cref{thm:abab-outerstring}, we can transfer known hardness results for outerstring graphs to \abab-free ordered graphs.
In particular, Unger~\cite{DBLP:conf/stacs/Unger88} proved that for every fixed $q \geq 4$, the \col{$q$} problem is \NP-hard in circle graphs, which form a subclass of outerstring graphs.
This immediately yields the following result.

\thmababhard*

\section[Excluding two nested edges or their extension]{Excluding \abba or its extensions}

\subsection{Algorithms}
The goal of this section is to show the following result.

\thmabbaalg*

We will use a standard decomposition of \akbbka-free graphs given in the following lemma; see Claims 3.11.1 and 3.11.2 in \cite{DBLP:journals/corr/abs-2604-24343}.

\begin{lemma}
    \label{lem:abbadecomposition}
    Let $k\geq 0$ be a fixed integer and let $G$ be a connected \akbbka-free graph with at least $2$ vertices. In polynomial time one can compute a sequence of pairs $((Z_i, R_i))_{i=1}^{t}$ for some $t\geq 1$ such that: 
    \begin{enumerate}
        \item the sets $Z_1, Z_2, \ldots, Z_t$ form a partition of $V(G)$ into consecutive segments, i.e. $Z_1\prec \ldots \prec Z_t$,
        \item every edge of $G$ in either contained in one of the sets $Z_i$ or connects two consecutive sets,
        \item for each $i\in [t]$, we have $\left| R_i \right| \leq 2k+2$ and the set $Z_i \setminus N[R_i]$ is independent.
    \end{enumerate}
\end{lemma}

The following definition captures what is left of the graph after applying Lemma \ref{lem:abbadecomposition} and removing the closed neighborhood of all the $R_i$'s.

A $k$-\emph{snake} is a \akbbka-free ordered graph $G$ together with a partition of its vertex set $V(G)=\bigcup_{i=\ell}^{\ell'} V_i$, into \emph{layers} so that:
    \begin{itemize}
        \item each layer is a segment;
        \item the only edges of $G$ join vertices from two consecutive layers.
    \end{itemize}
    The number of layers is called the \emph{length} of the snake.

A key part of our proof is the following lemma that $k$-snakes have separator dominated by a relatively small set.

\begin{restatable}{lemma}{lemsnakeseparator}\label{lem:snake_separator}
In a $k$-snake $G$ of length $\ell$, there exists a $1/6$-balanced separator dominated by a set of size at most $4(k+1)\ell$.
\end{restatable}

Before proving Lemma \ref{lem:snake_separator} let us show how it can be used to obtain a small separator in an \abbak-free graph with low maximum degree.

\begin{lemma}\label{lem:abba_separator}
Let $k\geq 0$ be a fixed integer. Every \abbak-free graph with at most $n$ vertices and maximum degree at most $n^{1/3}$ contains a $1/6$-balanced separator separator of order $\Oh(n^{2/3})$.
\end{lemma}
\begin{proof}
Without loss of generality, we may assume that $G$ is \akbbka-free. Indeed, consider an independent set $I_b$ of size $k$ in $G$ that is selected greedily, starting with the first vertex and traversing other vertices in the same order as they appear in $G$, until $k$ vertices are selected; and an independent set $I_e$ that is also selected greedily, but starting with the last vertex of $G$ and traversing vertices in reverse order. Note that if those procedures do not result in independent sets of size $k$ or it is not true that $I_b\prec I_e$, then $G$ has a dominating set of size smaller than $2k$, and together with the assumption on maximum degree of $G$ it implies that the order of $G$ is a constant. On the other hand, if $I_b\prec I_e$ and both sets are of size $k$, then the graph $G'=G\setminus (N[I_b] \cup N[I_e])$ is \akbbka-free; so if we find a desired separator $S'$ in $G'$, then $S'\cup N[I_b] \cup N[I_e]$ is a desired separator in $G$ of the same size, up to a lower-order term.

Note that if $G$ is not connected, then it is enough to find a separator in the largest connected component of $G$, so without loss of generality we assume that $G$ is connected. 

Now we apply Lemma \ref{lem:abbadecomposition} to $G$, but shift all the indices so that $Z_0$ contains the middle vertex of $G$. Therefore, we have a partition of $V(G)$ into sets $(Z_i)_{i=-s}^{i=t}$ such that each $Z_i$ is a segment in $G$ and there are no edges in $G$ between any two nonconsecutive sets; and a sequence of sets $(R_i)_{i=-s}^{i=t}$ such that $\left| R_i \right|\leq 2k+2$ and $Z_i \setminus R_i$ is independent for every $i$. 

We will argue that without loss of generality we may assume that $s+t+1$ is at most $n^{1/3}$. If the size of $Z_0$ is smaller than $n^{2/3}$, then $Z_0$ is the desired balanced separator. Otherwise, take $a$ and $b$ such that $Z_a$ is the last part with negative index of size smaller than $n^{2/3}$ and $Z_b$ is the first part with positive index of size smaller than $n^{2/3}$ (and set $a=-\infty$ or $b=\infty$ if the respective parts are undefined); and consider the graph $G'$ obtained from $G$ by removing all $Z_i$'s with $i\leq a$ and $i\geq b$. If we construct a balanced separator $S'$ in $G'$ of order at most $\Oh(n^{2/3})$, then $S'\cup Z_a \cup Z_b$ is a balanced separator in $G$ of the same order. Therefore, we only need to consider the case where all parts have size at least $n^{2/3}$, so clearly there are at most $n^{1/3}$ of them.

Now consider the graph $G'=G\setminus N[\bigcup R_i]$. Note that $G'$ is a $k$-snake of length at most $n^{1/3}$, so by Lemma \ref{lem:snake_separator} there is a $1/6$-balanced separator $S'$ in $G'$ dominated by a set of size at most $4(k+1)n^{1/3}$. Therefore, the set $S:=N[\bigcup R_i] \cup S'$ is a $1/6$-balanced separator in $G$ and, by the assumption on the maximum degree in $G$, the size of $S$ is at most $\Oh(n^{2/3})$, which completes the proof.
\end{proof}

Now we will prove Lemma \ref{lem:snake_separator}.

\lemsnakeseparator*

\begin{proof}[Proof of Lemma \ref{lem:snake_separator}]

    Let us begin by introducing some notation. 
    Let $n=|V(G)|$.
    Let $V_0$ be the layer containing the middle vertex.
    We call it the \emph{central} layer.
    For a vertex $v$, let $V(v)$ be the layer containing it.
    For a vertex $v$, let $\up (v)=\{u: u\succ v, u \in V(v)\}$ and $\down (v)=\{u: u\prec v, u \in V(v)\}$.
    In other words, these are the sets of vertices succeeding and preceding $v$ within its layer.
    Moreover, we shall write $V_{\geq t}=\bigcup_{i\geq t} V_i$ and similarly for $\leq$.

    We construct the set $P_\downarrow (v)$ inductively, as follows.

    We start by putting $v=v_0^0$ and greedily choosing $k$ independent vertices following $v$ to the set $W_0$. 
    Next, at each layer $V_i$, we extend $P_\downarrow (v)$ by including the following vertices.
    \begin{itemize}
        \item Pick $v_{i+1}^0 = \min \{N(\up(v_i^0)\setminus W_i)\cap V_{i+1}\}$;
        \item greedily choose $k$ vertices following $v_{i+1}^0$, independent from each other and $P_\downarrow (v)$, forming the set $W_{i+1}$;
        \item likewise, choose $k$ following $v_{i}^0$, independent from each other and $P_\downarrow (v)$, to form the set $W'_i$
        \item put $v_{i+1}^1 = \min \{N(v_i^0)\setminus W_{i+1}\cap V_{i+1}\}$;
        \item set $v_i^2 = \min \{N(v_{i+1}^0)\setminus W'_i\cap \up(v_i^0)\}$.        
    \end{itemize}
    We do this for each layer $V_0, V_1, V_2, \ldots$, and again symmetrically for layers $V_0, V_{-1}, V_{-2}, \ldots$.
    If at any point we cannot find the next vertex, simply terminate.
    Next, for $v\in V_0$ put $\Up(v)=\bigcup_i \up(v_i^0)$ and $\overline\Up(v)=G[V(G)\setminus \Up(v)]$.
    We claim that this set's neighborhood separates the two.

    \begin{claim}
        The set $N[P_\downarrow(v)]$ separates $\Up(v)$ from $\overline\Up(v)$.
    \end{claim}
    \begin{claimproof}
        Suppose there's an edge $uu'$ with $u\in \Up(v)\setminus N[P_\downarrow(v)]$, $u'\in \overline\Up(v) \setminus N[P_\downarrow(v)]$.
        Let $u\in V_i$ and assume, without loss of generality, that $i\geq 0$ and $u'\in V_{i+1}$.
        It follows that $u \succ W'_i \succ v_i^0$ and $u\prec v_{i+1}^0\prec W_{i+1} \prec v_{i+1}^1$.
        Thus, $\{u, u', v_i^0, v_{i+1}^1\}$, together with $W'_i$ and $W_{i+1}$ induce an \akbbka, a contradiction.
    \end{claimproof}

    Next, we claim that the size of $\Up(v)$ is monotonic with respect to $v$.

    \begin{claim}
        If $v, v' \in V_0$ and $v\prec v'$, then $\Up(v) \supseteq \Up(v')$. In particular, $|\Up(v)|\geq |\Up(v')|$.
    \end{claim}
    \begin{claimproof}
        It follows from the construction of $P_\downarrow(v)$ and the fact that if $X \subset X'$, then $N[X]\subset N[X']$.
    \end{claimproof}

    Finally, let $\Top=\{v\in V_0: |\Up(v)|\leq\frac{1}{6}n\}$.
    From the last claim it follows, that $\Top$ is a segment.

    Now, repeat the construction of $P_\downarrow(v)$ with $\min$ replaced by $\max$ and $\up$ by $\down$ to obtain the set $P_\uparrow(v)$ and sets $\Down(v), \overline\Down(v), \Bottom$.
    It is straightforward to verify that the analogues of all our observations so far hold for $P_\uparrow(v)$ and sets $\Down(v), \overline\Down(v), \Bottom$.

    Observe also that $V_0\subset \Up(v)\cup \Down(v)$ for any $v\in V_0$.
    
    As we already know that $N[P_\downarrow(v)]$, $N[P_\uparrow(v)]$ are separators; what remains is to combine them so that the separation is balanced.

    \paragraph{Case 1: $V_0 \ne \Top\cup \Bottom$.} Take $v\in V_0\setminus \Top \setminus \Bottom$.
    By definition, $|\Up(v)|, |\Down(v)| \geq \frac{1}{6}n$.
    Therefore, if neither $N[P_\downarrow(v)]$ nor $N[P_\uparrow(v)]$ is a $1/6$-balanced separator, it means that $|\overline\Up(v)|, |\overline\Down(v)| < \frac{1}{6}n$.
    In that case, $|\Up(v) \cap \Down(v) \cap V_{\geq 0}|, |\Up(v) \cap \Down(v) \cap V_{\leq 0}| \geq (\frac{1}{2}-2\cdot \frac{1}{6})\cdot n=\frac{1}{6}n$.
    As these sets are separated by $N[P_\downarrow(v) \cup P_\uparrow (v)]$, this finishes the proof in this case.
    
    \paragraph{Case 2: $V_0=\Top\cup \Bottom$.} Let $v_\mathsf{T}=\min(\Top)$ and $v_\mathsf{B}=\max(\Bottom)$. 
    Because $v_\mathsf{T} \prec v_\mathsf{B}$, it follows that $N[P_\downarrow(v_\mathsf{T})\cup P_\uparrow(v_\mathsf{B})]$ separates $|\overline\Up(v_\mathsf{T}) \cap \overline\Down(v_\mathsf{B}) \cap V_{\geq 0}| $ from $ |\overline\Up(v_\mathsf{T}) \cap \overline\Down(v_\mathsf{B}) \cap V_{\leq 0}|$, each of size at least $\frac{1}{6}n$.

    The proof is complete.
    
\end{proof}

Now we are ready to prove Theorem \ref{thm:abba-algorithms}. Having established Lemma \ref{lem:abba_separator}, we can get the desired running time with a standard branching algorithm.

\begin{proof}[Proof of Theorem \ref{thm:abba-algorithms}] 
We will construct an algorithm that, given an instance $I=(G, L)$ of \lcol{3} with \abbak-free graph $G$ where the sum the sizes of all the lists is at most $s$, will either accept of reject $I$.

We start by exhaustively applying two following reductions. If there is a vertex $v$ such that the list of $v$ contains exactly one possible color $c$, we remove $c$ from the lists of all neighbors of $v$ and then remove $v$ from the instance. If there is a vertex $v$ such that the list of $v$ is empty, we immediately reject the instance, and if the instance has no vertices, we immediately accept it.

If applying the above reductions leaves an instance with at least one vertex of degree at least $s^{1/3}$, we pick one such vertex, call it $v$, and for every color $c$ on the list of $v$ we produce an instance $I_c$ that is obtained from $I$ by coloring $v$ with the color $c$ (technically, it is replacing the list of $v$ with $\lbrace c\rbrace$). We the recursively solve those instances and if any of them was accepted, we accept $I$, otherwise we reject $I$.

In the remaining case, when all vertices in $I$ have degrees smaller than $s^{1/3}$, we take $S$ to be the smallest $\frac{1}{6}$-balanced separator in $I$. For every coloring $C$ of $S$, such that the color of each vertex $v\in S$ is from the list of $v$, we produced two instances $I_C^1$ and $I_C^2$ of the problem by coloring vertices of $S$ according to $C$ and splitting the connected components of $G\setminus S$ as evenly as possible; to be more precise, we partition the connected components of $G\setminus S$ into two graphs $G_1, G_2$ such that none of them has more than $\frac{5}{6}v(G)$ vertices, fabricate list assignment $L'$ such that $L'$ is obtained from $L$ by removing $C(v)$ from the lists of all neighbors of $v$ for every vertex $v\in S$, and take $I_C^1=(G_1, L'|_{V(G_1)})$ and $I_C^2=(G_2, L'|_{V(G_2)})$. We then solve all those instances recursively and if for any coloring $C$ both $I_C^1$ and $I_C^2$ were accepted, we accept $I$, and otherwise we reject $I$.

Note the the algorithm is clearly correct, because whenever an instance $I$ is accepted, one can certify it with an $L$-coloring of $G$, and rejection of $I$ happens after we have verified that all the possible partial colorings (that assign color to either a vertex $v$ or all vertices from the set $S$) can't be extended to an $L$-coloring of $G$.

We will find an upper bound $T(s)$ on the running time of the algorithm on an instance $I=(G, L)$, where the sum the sizes of all the lists is at most $s$. For technical reasons we will need $T$ to be a non-decreasing convex function.

The initial reductions and verifying the maximum degree in the graph can be done in time $P(s)$ for some polynomial $P$. Now consider what happens if the algorithm branched on a single vertex $v$. The list of $v$ had $2$ or $3$ elements, so at most $3$ sub-instances are produced. Each neighbor $w$ of $v$ has a list of size at least $2$, so in at most one of the sub-instances the list of $w$ can remain unchanged, and in the remaining sub-instances it is smaller by one color. It follows the the largest sub-instance has sum all the list sizes at most $s-2$, while the remaining ones -- at most $s-\frac{1}{2}s^{1/3}$. Therefore, in order for $T(s)$ to be a proper upper bound on the running time, it needs to satisfy the following inequality.
$$
T(s)\leqslant T\left(s-2\right)+T\left(s-\frac{1}{2}s^{2/3}\right) + P(s).
$$

Now consider the case when the algorithm branches on a separator $S$. By Lemma \ref{lem:abba_separator} (applied with $n=s$), the size of $S$ is at most $\Oh(s^{2/3})$. Moreover, if we search for it by browsing all subsets of $V(G)$ in order of their size, the search will finish in time $2^{\Oh(s^{2/3}\log s)}$. Note that $S$ is a $\frac{1}{6}$-balanced separator in terms of vertex counts, so it is $\frac{2}{17}$-balanced in terms of the sum of the list sizes. If we assure that $T$ is a convex and non-decreasing upper bound on the running time, for every coloring $C$ of $S$ both sub-instances $I_C^1$ and $I_C^2$ can be solved in time $T\left(\frac{2}{17}s\right) + T\left( \frac{15}{17}s\right)$. Therefore, we need $T$ to satisfy the following inequality.

$$
T(s) \leqslant  3^{s^{2/3}}\left(T\left(\frac{2}{17}s\right) + T\left( \frac{15}{17}s\right)\right)+2^{\Oh(s^{2/3}\log s)}
$$

One can readily verify that there is a non-decreasing convex function $T(s)=2^{\Oh(s^{2/3}\log s)}$ that satisfies both inequalities and therefore is an upper bound on the running time of the algorithm. Since $s$ is a most three times the number of vertices of $G$, the proof of the theorem is complete.
\end{proof}

\subsection{Hardness}

In this section we prove that \lcol{$3$} is \NP-hard for \abba-free ordered graphs. Before we proceed to the proof,
we will need some auxiliary tools.
First, let us recall the following well-known observation whose proof we include for completeness.

\begin{lemma}\label{lem:listsonpath}
    Let $P$ be a path on at least five vertices and let $x,y$ be its endpoints.
    For every $i \in [3]$, there is a list assignment $L: V(P) \to 2^{[3]}$ such that $L(x) = L(y) = [3]$
    and:
    \begin{itemize}
        \item if $\phi$ is a proper coloring of $(P,L)$, then $(\phi(x),\phi(y)) \neq (i,i)$,
        \item every assignment $\phi: \{x,y\} \to [3]$ such that $(\phi(x),\phi(y)) \neq (i,i)$ can be extended to a proper coloring of $(P,L)$.
    \end{itemize}
\end{lemma}
\begin{proof}
    By symmetry, we may assume that $i=1$.
    Denote the consecutive vertices of $P$ by $v_1, \ldots, v_r$ for $r \geq 5$, where $v_1 = x$ and $v_r = y$.
    If $r$ is odd, we set the lists as follows:
    \[
    \begin{array}{c c c c c }
    v_1 & v_2, \ldots, v_{r-3} & v_{r-2} & v_{r-1} & v_r\\ \hline       
    \{1,2,3\} & \{1,2\} &  \{2,3\} & \{1,3\} & \{1,2,3\}    
    \end{array}
    \]
    If $r$ is even, we set the lists as follows:
    \[
    \begin{array}{c c c c c c}
    v_1 & v_2, \ldots, v_{r-4} & v_{r-3} & v_{r-2} & v_{r-1} & v_r\\ \hline       
    \{1,2,3\} & \{1,2\} &  \{2,3\} & \{2,3\} & \{1,2\} & \{1,2,3\}    
    \end{array}
    \]    
    In both cases, the first block of consecutive vertices with list $\{1,2\}$ has odd size, so the first vertex of this block ($v_2$) always has the same color as the last one ($v_{r-3}$ or $v_{r-4}$, depending on the case).
    It is straightforward to verify that the lists defined above satisfy the required properties.
\end{proof}

Second, we will use the following result of Dujmovi\'c and Wood~\cite[Theorem 20]{DBLP:journals/dmtcs/DujmovicW05}.

\begin{theorem}[Dujmovi\'c and Wood~\cite{DBLP:journals/dmtcs/DujmovicW05}]\label{thm:planarsubdivision}
    Every $n$-vertex planar graph $G$ has a subdivision $G'$ such that every edge of $G$ is subdivided at most $n-2$ times and $G'$ admits a vertex ordering such that the resulting ordered graph does not contain \abba as a subgraph.
    Furthermore, $G'$ and its ordering can be found in time polynomial in $n$.
\end{theorem}

Now we are ready to show the main result of this section.
\thmabbahard*
\begin{proof}
    We reduce from \col{$3$} on planar graphs, which is known to be \NP-hard~\cite{DBLP:journals/tcs/GareyJS76}.
    Let $G_1$ be a planar graph.
    Let $G_2$ be obtained from $G_1$ by replacing each edge $xy \in E(G_1)$
    by three internally vertex-disjoint paths between $x$ and $y$, each with 5 vertices, including $x$ and $y$.
    Clearly $G_2$ is planar.

    Now, we apply \cref{thm:planarsubdivision} to $G_2$ to obtain a subdivision $G_3$ of $G_2$ and an ordering of $G_3$ such that the resulting ordered graph does not contain \abba as a subgraph; in particular, it is \abba-free.
    We fix this ordering and treat $G_3$ as an ordered graph.
    
    The vertex set of $G_3$ consists of original vertices of $G_1$ and \emph{subdivision vertices} that were added during the construction of $G_2$ and $G_3$.
    Furthermore, we can decompose $G_3$ into a collection of internally vertex-disjoint paths, each with at least 5 vertices, such that the endpoints of each path are original vertices of $G_1$ and all internal vertices are subdivision vertices. Every such path corresponds to an edge of $G_1$ and there are exactly three such paths for every edge of $G_1$.

    For every edge $xy \in E(G_1)$ we proceed as follows.
    Let the paths in $G_3$ corresponding to $xy$ be $P_1, P_2, P_3$.
    For $i \in \{1,2,3\}$, we apply \cref{lem:listsonpath} to $P_i$ with $i$ as the distinguished color.
    Let $L$ be the list assignment of $G_3$ obtained in this way: every original vertex of $G_1$ receives the list $[3]$, and every subdivision vertex receives the list given by the application of \cref{lem:listsonpath} to the path containing it.
    Note that together these three paths force the vertices $x$ and $y$ to receive different colors in any proper coloring of $(G_3,L)$, while every pair of distinct colors is attainable.

    Consequently, $(G_3,L)$ is a yes-instance of \lcol{$3$} if and only if $G_1$ is a yes-instance of \col{$3$}.
    As the size of $G_3$ is polynomial in the size of $G_1$, and $(G_3,L)$ can be found in polynomial time,
    the proof is complete.
\end{proof}

\section[Excluding two aligned edges]{Excluding \aabb}
\thmaabbhard*
\begin{proof}
    We reduce from \lcol{$3$} on bipartite graphs, which is known to be \NP-hard~\cite{DBLP:journals/dam/JansenS97}.
    Let $(G,L)$ be an instance of this problem, where the bipartition of $G$ is $(X,Y)$.
    Now, we perform a standard reduction from \lcol{$3$} to \col{$3$}: we add a triangle on vertices $Q = \{q_1,q_2,q_3\}$ and connect a vertex $v \in V(G)$ to $q_i$ if and only if $i \notin L(v)$.
    It is straightforward to verify that the resulting graph $G'$ is 3-colorable if and only if $(G,L)$ is a yes-instance of \lcol{$3$}.

    We make $G'$ an ordered graph by fixing the following ordering of $V(G')$: we place the vertices of $X$ first, then the vertices of $Q$, and then the vertices of $Y$. The ordering within each set is arbitrary.
    For contradiction, suppose that $G'$ contains an induced copy of \aabb with consecutive vertices $a,b,c,d$.
    
    Note that $G'[X \cup Y]$ is \aabb-free, so at least one of the vertices $a,b,c,d$ belongs to $Q$.
    By mirror symmetry, suppose that $\{a,b\} \cap Q \neq \emptyset$.
    As $c,d$ are non-adjacent to $a,b$ and thus cannot be in $Q$, we conclude that $c,d \in Y$.
    However, this is a contradiction as $Y$ is independent.
    This completes the proof.
\end{proof}
Next, we prove the following theorem.
\thmaabbalgo*

We will show the two statements of \cref{thm:aabb-algorithms} separately.

\begin{theorem}
     Given a \aabb-free ordered graph $G$, in polynomial time we can either decide that $\chi(G) > 3$, or find a proper $5$-coloring of $G$.
\end{theorem}
\begin{proof}
    If $G$ is edgeless, then it is trivially $1$-colorable and we are done.
    Thus, let $xy$, where $x \prec y$, be the edge of $G$ chosen so that $y$ is smallest possible with respect to $\prec$.
    Define the following sets of vertices:
    \begin{align*}
        A = & \ N(\{x,y\}) \\
        B = & \ \{v \notin A \cup \{x,y\} \mid v \prec y\} \\
        C = & \ \{v \notin A \cup \{x,y\} \mid v \succ y\}.
    \end{align*}
    Note that sets $\{x,y\}, A, B, C$ form a partition of $V(G)$.
    Furthermore, $B$ is independent by the choice of $xy$ and $C$ is independent as $G$ is \aabb-free.
    
    Color $x$ with color $1$ and $y$ with color $2$.
    Create an instance of \lcol{$3$} on $G[A]$ by first assigning the list $[3]$ to every vertex $v \in A$,
    and then removing a color $1$ or $2$ (or both) depending on whether $v$ is adjacent to $x$ or $y$ (or both).
    Note that since $A = N(\{x,y\})$, every vertex of $A$ has list of size at most $2$.
    Consequently, the instance $(G[A],L)$ can be solved in polynomial time using the algorithm from \cref{thm:edwards}.

    If the algorithm reports that $(G[A],L)$ is a no-instance, then we conclude that $\chi(G) > 3$ and we are done (note that the coloring of $x$ and $y$ with colors $1$ and $2$ was safe by symmetry).
    Otherwise, we obtain a proper coloring of $G[A]$ using colors $1,2,3$ that is consistent with the coloring of $x,y$.
    Finally, we color all vertices of $B$ with color $4$ and all vertices of $C$ with color $5$.
    This completes the proof.
\end{proof}

In the next result, we will use two observations from~\cite{DBLP:conf/esa/BielinskiPR26}.
For an ordered graph $G$, define 
\[
    \mathsf{Seagulls}(G)=\{ \{x,y,z\} \ | \ x\prec y\prec z \ \land \ xy\in E(G) \ \land \ yz\in E(G)\}.
\]

\begin{lemma}\label{lem:seagullsbipartite}
    Let $G$ be an ordered graph.
    If $\mathsf{Seagulls}(G) = \emptyset$, then $G$ is bipartite.
\end{lemma}

\begin{lemma}\label{lem:seagullsaabb}
    Let $G$ be an ordered graph.
    If $\mathsf{Seagulls}(G)$ contains two disjoint, non-adjacent sets,
    then $G$ contains an induced copy of \aabb.
\end{lemma}

Now we are ready to prove the second statement of \cref{thm:aabb-algorithms}.

\begin{theorem}
     Given an $n$-vertex \aabb-free ordered graph $G$, in time $n^{\Oh(\log n)}$ we can either decide that $\chi(G) > 3$, or find a proper $4$-coloring of $G$.
\end{theorem}
\begin{proof}
It will be convenient to work with a more general \lcol{$3$} problem.
For a given instance $G$ of \col{$3$}, we turn it into an instance $(G,L)$ of \lcol{$3$} by setting list $L(v) = [3]$ for every $v\in V(G)$.
Clearly, $(G,L)$ is a yes-instance of \lcol{$3$} if and only if $G$ is $3$-colorable.

We aim to use \cref{lem:branching} for $t=1$, $d=3$, $\eps=1/3$, and the obstacle family $\cF = \mathsf{Seagulls}(G)$.
Let us verify the assumptions of \cref{lem:branching}.
It suffices to show that for every induced subgraph $G'$ of $G$, if $\cF|_{V(G')} \neq \emptyset$, then there is a vertex $v \in \bigcup_{F \in \cF|_{V(G')}} F$ such that the set $N_{G'}[v]$ intersects at least $1/3$ of the sets in $\cF|_{V(G')}$.
So let $\{x,y,z\}\in \cF|_{V(G')}$.
By \cref{lem:seagullsaabb}, the set $N_{G'}[\{x,y,z\}]$ intersects every set in $\cF|_{V(G')}$.
Thus, by the pigeonhole principle, there is $v\in \{x,y,z\}$ such that $N_{G'}[v]$ intersects at least $1/3$ of the sets in $\cF|_{V(G')}$ as desired.

We call \cref{lem:branching} and in time $n^{\Oh(\log n)}$ we either
\begin{enumerate}[(i)]
\item conclude that $(G,L)$ is a no-instance of \lcol{$3$}, or
\item construct a family $\cG$ of $n^{\Oh(\log n)}$ triples $(G',L',\phi')$ as in \cref{lem:branching}.
\end{enumerate}

In case (i) we can conclude that $\chi(G)>3$.
So now consider case (ii). 
If for every $(G',L',\phi')\in \cG$, there is $v\in V(G')$ such that $L'(v)=\emptyset$, then each such $(G',L')$ is a no-instance of \lcol{$3$}.
By \cref{lem:branching}~(3), it means that $(G,L)$ is a no-instance of \lcol{$3$}, and this in turn implies that $\chi(G)>3$.

So now assume that there is a triple $(G',L',\phi')\in \cG$ such that for every $v\in V(G')$, it holds that $L'(v)\neq\emptyset$.
By \cref{lem:branching}~(2), we have that $\cF|_{V(G')}=\emptyset$, and therefore, by \cref{lem:seagullsbipartite}, $G'$ is bipartite.
Let $X,Y$ be the bipartition classes of $G'$ -- we can find them in polynomial time.
We define a proper $4$-coloring $\phi$ of $G$ as follows.
For every $v\in V(G)\setminus V(G')$, we set $\phi(v)=\phi'(v)$.
Furthermore, for every $v\in X$, we fix $\phi(v)$ as an arbitrary color from $L'(v)$; it exists as $L'(v)\neq \emptyset$.
Recall that $X$ is independent and by \cref{lem:branching}~(1), for any edge $uv\in E(G)$ with $u\in V(G)\setminus V(G')$, $v\in X$, we have $\phi(u)=\phi'(u)\notin L'(v)$, so the partial coloring $\phi$ obtained so far is proper.
Finally, we set $\phi(v)=4$ for every $v\in Y$. 
Since $Y$ is independent and the color $4$ is only used on $Y$, we obtain a proper $4$-coloring of $G$. This completes the proof.
\end{proof}

\subparagraph{AI disclosure.} ChatGPT was to simplify some arguments and to identify or correct omissions and minor mistakes. The authors take full responsibility for the content of the paper.

\bibliographystyle{plain}
\bibliography{main}

@inproceedings{DBLP:conf/stacs/Unger88,
  author       = {Walter Unger},
  editor       = {Robert Cori and
                  Martin Wirsing},
  title        = {On the k-Colouring of Circle-Graphs},
  booktitle    = {{STACS} 88, 5th Annual Symposium on Theoretical Aspects of Computer
                  Science, Bordeaux, France, February 11-13, 1988, Proceedings},
  series       = {Lecture Notes in Computer Science},
  volume       = {294},
  pages        = {61--72},
  publisher    = {Springer},
  year         = {1988},
  url          = {https://doi.org/10.1007/BFb0035832},
  doi          = {10.1007/BFB0035832},
  bibsource    = {dblp computer science bibliography, https://dblp.org}
}

@article{DBLP:journals/dmtcs/DujmovicW05,
  author       = {Vida Dujmovic and
                  David R. Wood},
  title        = {Stacks, Queues and Tracks: Layouts of Graph Subdivisions},
  journal      = {Discret. Math. Theor. Comput. Sci.},
  volume       = {7},
  number       = {1},
  pages        = {155--202},
  year         = {2005},
  url          = {https://doi.org/10.46298/dmtcs.346},
  doi          = {10.46298/DMTCS.346},
  bibsource    = {dblp computer science bibliography, https://dblp.org}
}

@article{DBLP:journals/iandc/GolovachPS14,
  author       = {Petr A. Golovach and
                  Dani{\"{e}}l Paulusma and
                  Jian Song},
  title        = {Closing complexity gaps for coloring problems on {H}-free graphs},
  journal      = {Inf. Comput.},
  volume       = {237},
  pages        = {204--214},
  year         = {2014},
  url          = {https://doi.org/10.1016/j.ic.2014.02.004},
  doi          = {10.1016/J.IC.2014.02.004},
}

@article{DBLP:journals/siamdm/HajebiLS24,
  author       = {Sepehr Hajebi and
                  Yanjia Li and
                  Sophie Spirkl},
  title        = {{List-3-Coloring} Ordered Graphs with a Forbidden Induced Subgraph},
  journal      = {{SIAM} J. Discret. Math.},
  volume       = {38},
  number       = {1},
  pages        = {1158--1190},
  year         = {2024},
  url          = {https://doi.org/10.1137/22m1515768},
  doi          = {10.1137/22M1515768},
  bibsource    = {dblp computer science bibliography, https://dblp.org}
}

@article{DBLP:journals/algorithmica/BonnetR19,
  author       = {{\'{E}}douard Bonnet and
                  Pawe\l{} Rz\k{a}\.zewski},
  title        = {Optimality Program in Segment and String Graphs},
  journal      = {Algorithmica},
  volume       = {81},
  number       = {7},
  pages        = {3047--3073},
  year         = {2019},
  url          = {https://doi.org/10.1007/s00453-019-00568-7},
  doi          = {10.1007/S00453-019-00568-7},
  bibsource    = {dblp computer science bibliography, https://dblp.org}
}

@inproceedings{DBLP:conf/sosa/PilipczukPR21,
  author       = {Marcin Pilipczuk and
                  Micha\l{} Pilipczuk and
                  Pawe\l{} Rz\k{a}\.ewski},
  editor       = {Hung Viet Le and
                  Valerie King},
  title        = {Quasi-polynomial-time algorithm for Independent Set in {$P_t$}-free
                  graphs via shrinking the space of induced paths},
  booktitle    = {4th Symposium on Simplicity in Algorithms, {SOSA} 2021, Virtual Conference,
                  January 11-12, 2021},
  pages        = {204--209},
  publisher    = {{SIAM}},
  year         = {2021},
  url          = {https://doi.org/10.1137/1.9781611976496.23},
  doi          = {10.1137/1.9781611976496.23},
  bibsource    = {dblp computer science bibliography, https://dblp.org}
}

@article{DBLP:journals/corr/abs-2505-00412,
  author       = {Esther Galby and
                  Paloma T. Lima and
                  Andrea Munaro and
                  Amir Nikabadi},
  title        = {Maximum list $r$-colorable induced subgraphs in {$kP_3$}-free
                  graphs},
  journal      = {CoRR},
  volume       = {abs/2505.00412},
  year         = {2025},
  url          = {https://doi.org/10.48550/arXiv.2505.00412},
  doi          = {10.48550/ARXIV.2505.00412},
  eprinttype    = {arXiv},
  eprint       = {2505.00412},
  bibsource    = {dblp computer science bibliography, https://dblp.org}
}

@article{DBLP:journals/combinatorica/ChudnovskyHS24,
  author       = {Maria Chudnovsky and
                  Sepehr Hajebi and
                  Sophie Spirkl},
  title        = {{List-$k$-Coloring $H$-Free Graphs for All $k > 4$}},
  journal      = {Comb.},
  volume       = {44},
  number       = {5},
  pages        = {1063--1068},
  year         = {2024},
  url          = {https://doi.org/10.1007/s00493-024-00106-2},
  doi          = {10.1007/S00493-024-00106-2},
  bibsource    = {dblp computer science bibliography, https://dblp.org}
}

@article{DBLP:journals/siamdm/HajebiLS22,
  author       = {Sepehr Hajebi and
                  Yanjia Li and
                  Sophie Spirkl},
  title        = {Complexity Dichotomy for List-5-Coloring with a Forbidden Induced
                  Subgraph},
  journal      = {{SIAM} J. Discret. Math.},
  volume       = {36},
  number       = {3},
  pages        = {2004--2027},
  year         = {2022},
  url          = {https://doi.org/10.1137/21m1443352},
  doi          = {10.1137/21M1443352},
  bibsource    = {dblp computer science bibliography, https://dblp.org}
}

@article{DBLP:journals/algorithmica/0001GKP15,
  author       = {Jean{-}Fran{\c{c}}ois Couturier and
                  Petr A. Golovach and
                  Dieter Kratsch and
                  Dani{\"{e}}l Paulusma},
  title        = {List Coloring in the Absence of a Linear Forest},
  journal      = {Algorithmica},
  volume       = {71},
  number       = {1},
  pages        = {21--35},
  year         = {2015},
  url          = {https://doi.org/10.1007/s00453-013-9777-0},
  doi          = {10.1007/S00453-013-9777-0},
  bibsource    = {dblp computer science bibliography, https://dblp.org}
}

@article{DBLP:journals/algorithmica/HoangKLSS10,
  author       = {Ch{\'{\i}}nh T. Ho{\`{a}}ng and
                  Marcin Kami\'nski and
                  Vadim V. Lozin and
                  Joe Sawada and
                  Xiao Shu},
  title        = {Deciding \emph{k}-Colorability of {$P_5$}-Free Graphs
                  in Polynomial Time},
  journal      = {Algorithmica},
  volume       = {57},
  number       = {1},
  pages        = {74--81},
  year         = {2010},
  url          = {https://doi.org/10.1007/s00453-008-9197-8},
  doi          = {10.1007/S00453-008-9197-8},
  bibsource    = {dblp computer science bibliography, https://dblp.org}
}

@article{DBLP:journals/ejc/Huang16,
  author       = {Shenwei Huang},
  title        = {Improved complexity results on $k$-coloring {$P_t$}-free graphs},
  journal      = {Eur. J. Comb.},
  volume       = {51},
  pages        = {336--346},
  year         = {2016},
  url          = {https://doi.org/10.1016/j.ejc.2015.06.005},
  doi          = {10.1016/J.EJC.2015.06.005},
  bibsource    = {dblp computer science bibliography, https://dblp.org}
}

@article{DBLP:journals/siamcomp/Holyer81a,
  author       = {Ian Holyer},
  title        = {The {NP}-Completeness of Edge-Coloring},
  journal      = {{SIAM} J. Comput.},
  volume       = {10},
  number       = {4},
  pages        = {718--720},
  year         = {1981},
  url          = {https://doi.org/10.1137/0210055},
  doi          = {10.1137/0210055},
  bibsource    = {dblp computer science bibliography, https://dblp.org}
}

@article{DBLP:journals/cpc/Emden-WeinertHK98,
  author       = {Thomas Emden{-}Weinert and
                  Stefan Hougardy and
                  Bernd Kreuter},
  title        = {Uniquely Colourable Graphs and the Hardness of Colouring Graphs of
                  Large Girth},
  journal      = {Comb. Probab. Comput.},
  volume       = {7},
  number       = {4},
  pages        = {375--386},
  year         = {1998},
  url          = {http://journals.cambridge.org/action/displayAbstract?aid=46667},
  bibsource    = {dblp computer science bibliography, https://dblp.org}
}

@article{DBLP:journals/jal/LevenG83,
  author       = {Daniel Leven and
                  Zvi Galil},
  title        = {{NP} Completeness of Finding the Chromatic Index of Regular Graphs},
  journal      = {J. Algorithms},
  volume       = {4},
  number       = {1},
  pages        = {35--44},
  year         = {1983},
  url          = {https://doi.org/10.1016/0196-6774(83)90032-9},
  doi          = {10.1016/0196-6774(83)90032-9},
  bibsource    = {dblp computer science bibliography, https://dblp.org}
}

@article{DBLP:journals/algorithmica/ChudnovskyHSZ21,
  author       = {Maria Chudnovsky and
                  Shenwei Huang and
                  Sophie Spirkl and
                  Mingxian Zhong},
  title        = {List 3-Coloring Graphs with No Induced {$P_6 + rP_3$}},
  journal      = {Algorithmica},
  volume       = {83},
  number       = {1},
  pages        = {216--251},
  year         = {2021},
  url          = {https://doi.org/10.1007/s00453-020-00754-y},
  doi          = {10.1007/S00453-020-00754-Y},
  bibsource    = {dblp computer science bibliography, https://dblp.org}
}

@article{DBLP:journals/combinatorica/BonomoCMSSZ18,
  author       = {Flavia Bonomo and
                  Maria Chudnovsky and
                  Peter Maceli and
                  Oliver Schaudt and
                  Maya Stein and
                  Mingxian Zhong},
  title        = {Three-Coloring and List Three-Coloring of Graphs Without Induced Paths
                  on Seven Vertices},
  journal      = {Comb.},
  volume       = {38},
  number       = {4},
  pages        = {779--801},
  year         = {2018},
  url          = {https://doi.org/10.1007/s00493-017-3553-8},
  doi          = {10.1007/S00493-017-3553-8},
  bibsource    = {dblp computer science bibliography, https://dblp.org}
}

@article{DBLP:journals/tcs/Edwards86,
  author       = {Keith Edwards},
  title        = {The Complexity of Colouring Problems on Dense Graphs},
  journal      = {Theor. Comput. Sci.},
  volume       = {43},
  pages        = {337--343},
  year         = {1986},
  url          = {https://doi.org/10.1016/0304-3975(86)90184-2},
  doi          = {10.1016/0304-3975(86)90184-2},
  bibsource    = {dblp computer science bibliography, https://dblp.org}
}

@article{DBLP:journals/corr/abs-2604-24343,
  author       = {Paweł Rafał Bieliński and
                  Marta Piecyk and
                  Paweł Rzążewski},
  title        = {Maximum Weight Independent Set in Hereditary Classes of Ordered Graphs},
  journal      = {CoRR},
  volume       = {abs/2604.24343},
  year         = {2026},
  url          = {https://doi.org/10.48550/arXiv.2604.24343},
  doi          = {10.48550/ARXIV.2604.24343},
  eprinttype   = {arXiv},
  eprint       = {2604.24343},
  bibsource    = {dblp computer science bibliography, https://dblp.org}
}

@inproceedings{DBLP:conf/stacs/PiecykR26,
  author       = {Marta Piecyk and
                  Pawel Rzazewski},
  editor       = {Meena Mahajan and
                  Florin Manea and
                  Annabelle McIver and
                  Kim Thang Nguyen},
  title        = {List Coloring Ordered Graphs with Forbidden Induced Subgraphs},
  booktitle    = {43rd International Symposium on Theoretical Aspects of Computer Science,
                  {STACS} 2026, Grenoble, France, March 9-13, 2026},
  series       = {LIPIcs},
  volume       = {364},
  pages        = {74:1--74:17},
  publisher    = {Schloss Dagstuhl - Leibniz-Zentrum f{\"{u}}r Informatik},
  year         = {2026},
  url          = {https://doi.org/10.4230/LIPIcs.STACS.2026.74},
  doi          = {10.4230/LIPICS.STACS.2026.74},
  bibsource    = {dblp computer science bibliography, https://dblp.org}
}

@article{DBLP:journals/tcs/GareyJS76,
  author       = {M. R. Garey and
                  David S. Johnson and
                  Larry J. Stockmeyer},
  title        = {Some Simplified NP-Complete Graph Problems},
  journal      = {Theor. Comput. Sci.},
  volume       = {1},
  number       = {3},
  pages        = {237--267},
  year         = {1976},
  url          = {https://doi.org/10.1016/0304-3975(76)90059-1},
  doi          = {10.1016/0304-3975(76)90059-1},
  bibsource    = {dblp computer science bibliography, https://dblp.org}
}

@article{DBLP:journals/dam/JansenS97,
  author       = {Klaus Jansen and
                  Petra Scheffler},
  title        = {Generalized Coloring for Tree-like Graphs},
  journal      = {Discret. Appl. Math.},
  volume       = {75},
  number       = {2},
  pages        = {135--155},
  year         = {1997},
  url          = {https://doi.org/10.1016/S0166-218X(96)00085-6},
  doi          = {10.1016/S0166-218X(96)00085-6},
  bibsource    = {dblp computer science bibliography, https://dblp.org}
}

@inproceedings{DBLP:conf/sosa/SGLS26,
  author       = {Ajaykrishnan E. S. and
                  Robert Ganian and
                  Daniel Lokshtanov and
                  Vaishali Surianarayanan},
  editor       = {Sepehr Assadi and
                  Eva Rotenberg},
  title        = {A Quasi-Polynomial Time Algorithm for 3-Coloring Circle Graphs},
  booktitle    = {2026 Symposium on Simplicity in Algorithms, {SOSA} 2026, Vancouver,
                  BC, Canada, January 12-14, 2026},
  pages        = {65--80},
  publisher    = {{SIAM}},
  year         = {2026},
  url          = {https://doi.org/10.1137/1.9781611978964.6},
  doi          = {10.1137/1.9781611978964.6},
  bibsource    = {dblp computer science bibliography, https://dblp.org}
}

@inproceedings{DBLP:conf/esa/BielinskiPR26,
  author       = {Pawel Rafal Bielinski and
                  Marta Piecyk and
                  Pawel Rzazewski},
  editor       = {Philip Bille and
                  Seth Pettie and
                  Sabine Storandt},
  title        = {Maximum Weight Independent Set in Hereditary Classes of Ordered Graphs},
  booktitle    = {34th Annual European Symposium on Algorithms, {ESA} 2026, L'Aquila,
                  Italy, August 31 - September 4, 2026},
  series       = {LIPIcs},
  volume       = {388},
  pages        = {30:1--30:15},
  publisher    = {Schloss Dagstuhl - Leibniz-Zentrum f{\"{u}}r Informatik},
  year         = {2026},
  url          = {https://doi.org/10.4230/LIPIcs.ESA.2026.30},
  doi          = {10.4230/LIPICS.ESA.2026.30},
  bibsource    = {dblp computer science bibliography, https://dblp.org}
}

@book{DBLP:books/sp/CyganFKLMPPS15,
  author       = {Marek Cygan and
                  Fedor V. Fomin and
                  Lukasz Kowalik and
                  Daniel Lokshtanov and
                  D{\'{a}}niel Marx and
                  Marcin Pilipczuk and
                  Michal Pilipczuk and
                  Saket Saurabh},
  title        = {Parameterized Algorithms},
  publisher    = {Springer},
  year         = {2015},
  url          = {https://doi.org/10.1007/978-3-319-21275-3},
  doi          = {10.1007/978-3-319-21275-3},
  isbn         = {978-3-319-21274-6},
  bibsource    = {dblp computer science bibliography, https://dblp.org}
}

\end{document}